\documentclass[12pt]{article}
 
\usepackage{url}
\usepackage{scicite}
\usepackage{cite}
\usepackage{amsmath}
\usepackage{amsfonts,bm}
\usepackage{amssymb}
\usepackage{graphicx}
\usepackage{float}
\usepackage{caption}
\usepackage{subcaption}

\newcommand{\st}{\text{subject to}\quad }

\newcommand{\norm}[1]{\left\lVert#1\right\rVert}

\newcommand\numberthis{\addtocounter{equation}{1}\tag{\theequation}} 

\DeclareMathOperator*{\minimize}{minimize}

\usepackage{xcolor}
\definecolor{orange}{RGB}{255,107,0}

\definecolor{green}{RGB}{0, 145, 17}

\usepackage{times}

\newenvironment{sciabstract}{%
\begin{quote} \bf}
{\end{quote}}

\newcounter{lastnote}
\newenvironment{scilastnote}{%
\setcounter{lastnote}{\value{enumiv}}%
\addtocounter{lastnote}{+1}%
\begin{list}%
{\arabic{lastnote}.}
{\setlength{\leftmargin}{.22in}}
{\setlength{\labelsep}{.5em}}}
{\end{list}}

\title{Uncovering migration systems through spatio-temporal tensor co-clustering}

\author
{Zack W. Almquist,$^{1\ast}$ Tri Duc Nguyen$^{2}$ Mikael Sorensen$^{3}$,\\ Xiao Fu$^{2}$, Nicholas D. Sidiropoulos$^{3}$\\
\\
\normalsize{$^{1}$Department of Sociology, Department of Statistics,}\\ 
\normalsize{Center for Studies in Demography and Ecology,}\\ 
\normalsize{Center for Statistics and Social Sciences, eScience,}\\ 
\normalsize{University of Washington, Seattle, WA, 9819}\\
\normalsize{$^{2}$Electrical Engineering and Computer Science, Oregon State University, OR, 97331}\\
\normalsize{$^{3}$Electrical and Computer Engineering, University of Virginia, VA, 22904}\\
\\
\normalsize{$^\ast$To whom correspondence should be addressed; E-mail:  zalmquist@uw.edu.}
}

\date{\today}

\begin{document} 


\baselineskip24pt


\maketitle

\clearpage

\begin{sciabstract}
\textbf{ABSTRACT}

A central problem in the study of human mobility is that of migration systems. Typically, migration systems are defined as a set of relatively stable movements of people between two or more locations over time. While these emergent systems are expected to vary over time, they ideally contain a stable underlying structure that could be discovered empirically. There have been some notable attempts to formally or informally define migration systems, however they have been limited by being hard to operationalize, and by defining migration systems in ways that ignore origin/destination aspects and/or fail to account for migration dynamics over time. In this work, we propose to employ \emph{spatio-temporal (ST) tensor co-clustering}---that stems from signal processing and machine learning theory---as a novel migration system analysis tool. 
Tensor co-clustering is designed to cluster entities exhibiting similar patterns across multiple modalities, and thus suits our purpose of analyzing spatial migration activities across time.
To demonstrate its effectiveness for describing stable migration systems, we first focus on domestic migration between counties in the US from 1990 to 2018.
We conduct three case studies, namely, (i) US Metropolitan Areas, (ii) the state of California, and (iii) Louisiana, in which the last focuses on detecting exogenous events such as Hurricane Katrina in 2005. 
In addition, we also look into  a case study at a larger scale, using world wide international migration data of 200 countries between 1990 and 2015. Finally, we conclude with discussion and limitations of this approach. 
\end{sciabstract}
\clearpage


\section*{Introduction}

A central problem to the study of migration is how to define and detect migration systems \cite{massey1999worlds,bakewell2014relaunching,kritz1992international,mabogunje1970systems,massey2015missing,yang2016comparative,girvan2002community}. Migration systems represent an ``emergent social entity,'' continually evolving and exchanging people over varying levels of spatial and temporal scales \cite{abel2021form}. There have been some notable attempts to formally or informally define migration systems \cite{abel2021form,expert2011uncovering}, but they have been limited by either being hard to operationalize \cite{dewaard2012migration} or by defining migration systems as symmetric (rather than directed origin/destination), static, or both. Most recently, the work by Abel et al.\cite{abel2021form} has employed a clustering algorithm (this is a major area of study in social network and network science, see for example \cite{clauset2005finding,boccaletti2007detecting,ahajjam2018new}) that allows for directed networks \cite{rosvall2008maps} to detect international migration systems over five year aggregates of migration data; however this work only considers clustering on static snapshots which are then strung together for analysis. This work, like previous research in the area leverages the idea that one can represent migration flows as a \emph{weighted-graph} or \emph{network} \cite{slater2009two,sorichetta2016mapping}. Here, we focus on ``raw'' migration data, i.e.,  simply the counts of individuals or households between two geographical units (e.g., United States and Mexico or Los Angeles County, CA and King County, WA). By representing the migration flows between such spatial units, one can employ tools from social networks  \cite{butts2009revisiting},  network science \cite{vespignani2018twenty,clauset2004finding} and other computational social science \cite{lazer2020computational} to analyze the data.


Social Network Analysis \cite{wasserman1994social} and Network Science \cite{barabasi2013network} have a long history of studying network clustering and  community detection problems. Classic community detection methods (see, e.g., \cite{airoldi2008mixed,karrer2011stochastic}) look for clusters of nodes in a graph or network. Major community detection methods in the literature include optimal modularity (e.g.,  \cite{good2010performance,fortunato2007resolution,lancichinetti2011limits}), edge-betweenness (e.g., \cite{girvan2002community}), leading eigenvector (e.g., \cite{newman2006finding}), fast-greedy (e.g., \cite{clauset2004finding,wakita2007finding}), multi-level (e.g., \cite{blondel2008fast}), walktrap (e.g., \cite{pons2005computing}), label propagation (e.g., \cite{raghavan2007near}), and infoMap (e.g., \cite{rosvall2009map}). These approaches typically do not take the temporal modality into consideration. However, incorporating the time domain information in community detection  is of interest to both social and physical sciences; for example, such techniques have been used to explain Biology mechanisms \cite{bonchev2007complexity,girvan2002community} and to group dynamics of Windsurfers \cite{almquist2014logistic}. 

Many network problems, such as international or domestic migration, are dynamic in nature and require a method which takes this into account. 
As such, there has been growing interest in applying community detection to dynamically evolving networks. This has largely been done through applying the classic community detection methods to network ``snapshots" or panel data and analyzing how the system has changed. For example, \cite{martinet2020robust} studies the {\it change} of node associations in graphs that are collected sequentially; and  \cite{nguyen2014dynamic} studies the computational aspects of adapting new community structures quickly based on previously estimated communities. A brief review of dynamic community detection methods can be found in the the book chapter \cite{cazabet2017dynamic}. 

Notably, the area of dynamic community detection (e.g.,  \cite{martinet2020robust,nguyen2014dynamic,cazabet2017dynamic}) focuses on change in community structure. Instead, our interest lies in discovering the {\it consistency} of community structure over time. Within the area of migration systems analysis there has been one attempt at looking at applying community detection methods to international migration The appraoch by Abel et al \cite{abel2021form} used the infoMap community detection method over five year migration flows, and subsequently analyzed the change in community structure over the observed time periods. In this article, we introduce a method for holistically measuring the community structure over time with a focus on the stable communities rather than differences. At the end of our results section we compare the international migration system in Abel et al to our own method. Further, we compare the walktrap method applied to pre- and post-Hurricane Katrina to our method where we find local clustering compared to a limited set of non-local clustering, and out method allows for overlapping clustering and a measure of significance for the community/migration system. 

In terms of methodology, we propose to employ a \emph{spatio-temporal (ST) tensor co-clustering} method from the signal processing and machine learning literature \cite{sidiropoulos2017tensor,papalexakis2013from}. 
Tensors are a natural format to store data having multiple modalities (e.g., the migration counts indexed by origin, destination, time). Tensors also encode the cross-modality dependencies using the notion of {\it tensor rank}, which is a high-order generalization of the matrix rank.
The ST tensor co-clustering method allows for a low tensor rank representation of a weighted spatial-temporal graph, e.g., origin-destination counts or other migration measures acquired over time. At each time point, the weighted graph is defined (in this case) by an origin-destination directed adjacency matrix where an edge represents the number of migrants from one spatial unit to another (e.g. Los Angeles county to New York county). This representation results in a data-driven migration system that meets the concept of a migration system in the literature (e.g. Massey et al. \cite[p. 61]{massey1999worlds}). That is, every \emph{rank-one} tensor extracted from the ST tensor co-clustering model represents a migration ``community'' (e.g., collection of counties) whose members maintain a spatial interaction pattern with each other and share a similar temporal profile.

The ST tensor co-clustering---under this data definition---identifies the stable temporal clusters of the weighted graph (e.g., migrant counts from United States and Mexico) and its temporal intensity over time
(e.g., the Mexican born population peaking in 2007 and decreasing post 2011\footnote{\tiny\url{https://www.pewresearch.org/hispanic/2012/04/23/ii-migration-between-the-u-s-and-mexico/}}; for attempts to estimate world migration rates; see \cite{azose2019estimation,abel2014quantifying}). To demonstrate the effectiveness of this approach, we consider two datasets and a number of case studies. First, we apply it to domestic migration data within the United States over the period of 1990 to 2018 and to international migration data at five year intervals between 1990 and 2015. The US Internal Revenue Service (IRS) makes publicly and freely available migration data at the state and county levels \cite{hauer2019irs,gross2005internal,pierce2015soi}\footnote{\url{https://www.irs.gov/statistics/soi-tax-stats-migration-data}}. These data are built from address information contained in year-to-year tax returns and covers approximately 87\% of all US households \cite{molloy2011internal}. The IRS migration data represents a particularly unique and valuable set of migration data for the US \cite{hauer2019irs}. In fact, the US Census Bureau uses the IRS migration data to produce state and county estimates of net-migration as part of its Population Estimate Program \cite{hauer2019irs}. This data set is ideally suited for testing ST co-clustering methods.
Our final case study is based around international migration. Specifically, we apply the ST tensor co-clustering approach to the international migration data constructed by Azose and Raftery \cite{azose2019estimation,abel2014quantifying} and updated by Abel et al. \cite{abel2021form}. International migration has a long history of theory on the migration systems with a strong interest in empirically finding stable country clusters over time, but with only limited actual methods and applications.  The ST tensor co-clustering method is again uniquely suited for this task.




\subsection*{Migration Systems}

Migration systems attempt to capture the persistent interchange of people between places over time \cite[p. 61]{massey1999worlds}, according to Massey et al. \cite[p. 61]{massey1999worlds}: "[t]he end result is a set of relatively stable exchanges of people between [places] ... yielding an identifiable geographic structure that persists across space and time." In particular, these systems are expected to be sustained over time, emergent and vary by spatial and temporal scales \cite{bakewell2014relaunching}, making them naturally representable by mathematical graphs or networks \cite{kritz1992international}. The expectation is that these systems should exist at the international and local levels \cite{abel2021form} as a hierarchical process. In this work we will look to operationalize this concept of a migration system. Through the ST tensor co-clustering algorithm, we aim to find stable spatio-temporal ``systems" in the United States internal migration and in the  international estimates of migration from 1990 to 2015 \cite{azose2019estimation}. 

\subsection*{Spatial-Temporal Tensor Co-clustering for Migration Systems}
The ST tensor co-clustering approach takes a three-way array as its input. The tensor has a size of $I \times I \times K$, where  $I$ is the number of geographical entities (e.g., counties, cities and countries) and $K$ is the number of temporal samples (e.g., years or months). 
The tensor is represented using the notation ${\cal X}\in\mathbb{R}^{I\times I\times K}$. 
Every entry of ${\cal X}$ has three coordinates. For example, in the IRS migration data that will be studied in this work, the entry ${\cal X}(i,j,k)$ represents the number of migrants moving from county $i$ to county $j$ in year $k$.
It can be regarded as a natural extension of a matrix whose entries only have two coordinates.
When fixing $k$, the matrix (or the $k$th ``tensor slab'') ${\cal X}(:,:,k)\in\mathbb{R}^{I\times I}$ is the weighted graph (e.g., origin to destination counts) collected in the $k$th year.
The diagonal entries of every such matrix are ignored during data analysis using an incomplete tensor decomposition technique. The reason is that the diagonal elements do not have meaning in the context of this migration flow analysis (i.e. we do not have measurements on within county or within country mobility patterns). 
The tensor co-clustering method decomposes ${\cal X}$ into the summation of $F$ rank-one tensors, where $F$ is pre-specified (we chose this based on information decay in the model fitting process).
After convergence of the co-clustering optimization algorithm,  $F$ migration communities (note that we use migration system and migration community in an exchangeable manner in this work) will be discovered.
Each community is represented by
a tuple of vectors $({\bf a}_f\in\mathbb{R}^I, {\bf b}_f\in\mathbb{R}^I, {\bf c}_f\in \mathbb{R}^K)$. 
The ${\bf a}_f$ vector is an origin entity indicator, where ${\bf a}_f(i)$ indicates the level of involvement of entity $i$ in the $f$th migration system. The ${\bf b}_f$ vector is defined in a similar way for destination entities. The vector ${\bf c}_f$ represents the temporal profile of the $f$th migration system, i.e., how active is this system in each year.
Furthermore, the matrix ${\bf a}_f {\bf b}_f^T $ represents the spatial association of origin and destination entities in the community, while $\bm{c}_f$ encodes temporal intensity of the association. 
The tuple forms a rank-one tensor $\mathcal{C}_f$ by the outer product operation, i.e.,
        \begin{align*}
            &\mathcal{C}_f = {\bf a}_f \circ {\bf b}_f \circ {\bf c}_f =({\bf a}_f{\bf b}_f^T)\circ {\bf c}_f^T \\
            &\mathcal{C}_f(i, i, k) =  0, \quad \text{for all $1 \leq i \leq I, 1\leq k \leq K$},
        \end{align*} 
where $\circ$ denotes the outer product; i.e.,
The above can also be expressed as $${\cal C}_f(i,j,k)={\bf a}_f(i){\bf b}_f(j){\bf c}_f(k).$$ The readers are referred to more detailed definitions of tensor operators in \cite{sidiropoulos2017tensor}.

\begin{figure}[H]
\centering
\includegraphics[width=.85\linewidth]{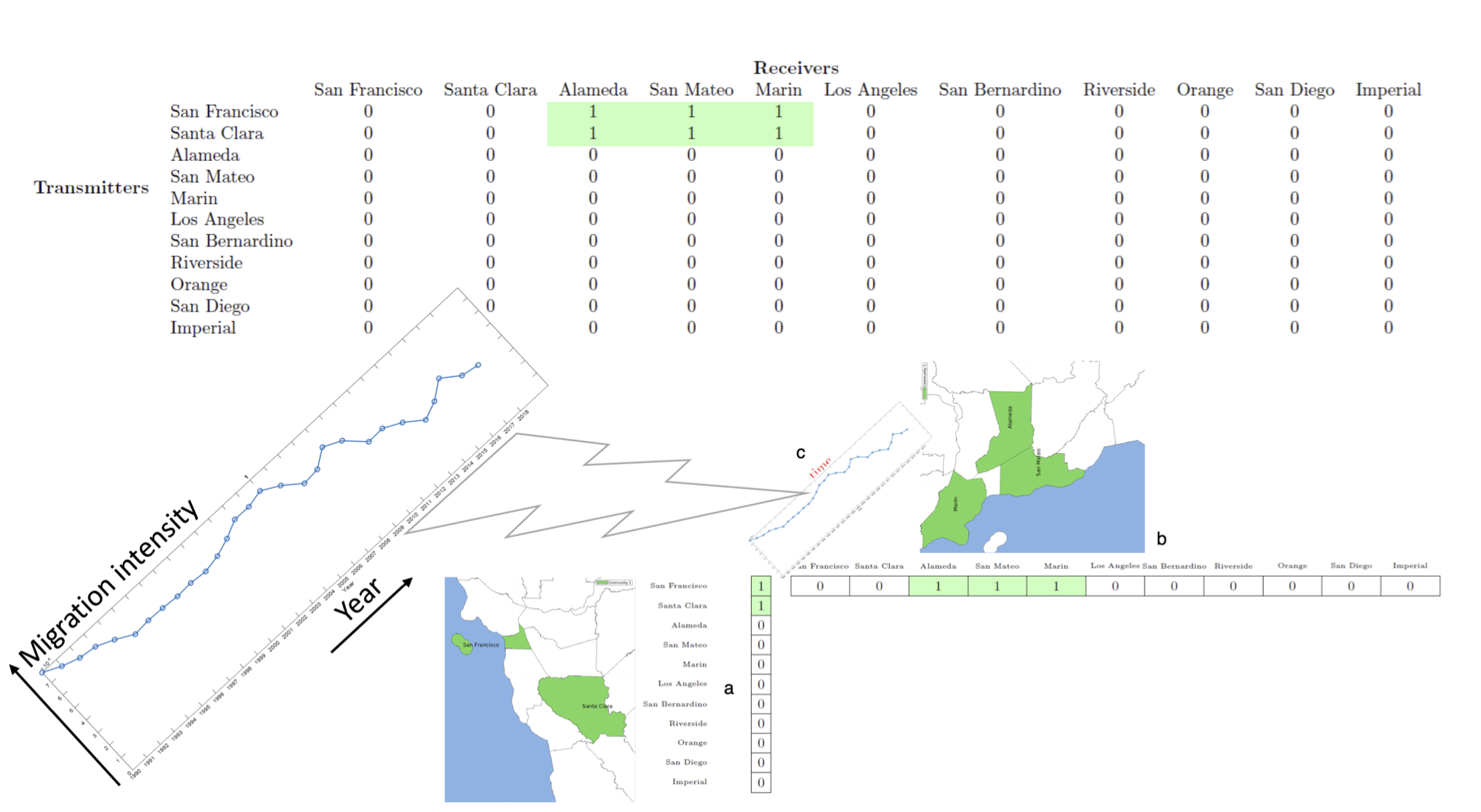}
\caption{An example of the rank-one tensor based representation of a stable migration system with its temporal profile. In this system, the origins are San Francisco and Santa Clara.  Hence, ${\bf a}(1)$ (San Francisco) and ${\bf a}(2)$ (Santa Clara) are nonzero. The destinations are Alameda, San Mateo, and Marin, and thus the corresponding ${\bf b}(j)$'s ($j=3,4,5$) are nonzero---as shown in the lower subfigure.
In addition, the top table shows ${\bf a} {\bf b}^T$, i.e., the spatial association of transmitters and receivers. The migration intensity is the ${\bf c}$ vector, which reflects how this system's activity level varies over the years. When multiple migration systems are simultaneously present, the associated data tensor is described by a sum of such spatio-temporal rank-one terms.}
\label{fig:tensorExample}
\end{figure}

This rank-one representation is exactly a stable migration system with time-varying activity levels. The rank-one tensor representation of a spatio-temporal migration system is illustrated in Figure~\ref{fig:tensorExample}.  In this system, the origins are San Francisco and Santa Clara.  Hence, ${\bf a}(1)$ (San Francisco) and ${\bf a}(2)$ (Santa Clara) are nonzero. The destinations are Alameda, San Mateo, and Marin, and thus the corresponding ${\bf b}(j)$'s ($j=3,4,5$) are nonzero---as shown in the lower subfigure.
In addition, the top table shows ${\bf a} {\bf b}^T$, i.e., the spatial association of transmitters and receivers. The migration intensity is the ${\bf c}$ vector, which reflects how this system's activity level varies over the years. When multiple migration systems are simultaneously present, the associated data tensor is described by a sum of such spatio-temporal rank-one terms.

\section*{Results}

To illustrate the viability of using the ST tensor co-clustering method for understanding migration systems, we focus on two datasets and four case studies: (i) US Metropolitan Areas, (ii) California, (iii) Louisiana with focus on detecting exogenous events such as Hurricane Katrina in 2005 and (iv) international migration from 1990 to 2015 at five year intervals. Case studies (i)-(iii) are conducted with an IRS migration data \cite{hauer2019irs} and case study (iv) is based on an international migration data from \cite{abel2014quantifying}.





\subsection*{IRS Data - US Census Metropolitan Statistical Areas}

Migration scholars have often focused on the economic, social and political impact of internal migration in the United States \cite{frey2017internal,greenwood1972determinants}. Internal migrants, unlike international migration, are attracted to destinations other than traditional port-of-entry. The origins and destinations can respond to ``pushes" and ``pulls" related to environmental, political and economic changes \cite{plane2005migration}.  Frey \cite{frey1996immigration} notes that metropolitan areas are more closely aligned with the concept of the labor market or community and potentially the most appropriate geographic units for examining internal migration patterns. Here, we can employ the ST tensor co-clustering algorithm to ``uncover" the stable migration systems over a time period. Further, we can observe the temporal change in migration intensity due to fluctuations (e.g., labor market) over the observed time period. 

The US Census Bureau defines 384 metropolitan statistical areas (MSAs) which represent one or more counties which contain at least one urbanized area of 50,000 or more inhabitants. There are 384 MSAs ranging in size from 20 Million (New York City-Newark-Jersey City) to 58,000 (Carson City, NV MSA). These 384 metropolitan areas represent 86\% of the US population in 2020 \cite{molloy2011internal}.  In this vein an important question is can we find migration systems between these major economic regions over the period 1990-2018? We follow up this question by asking how variable are these communities over time in the intensity of their activities. The full migration system decomposition can be found in the SI Appendix. Here we focus on just the first migration system, which is comprised of the major West Coast hub dominated by the Los Angeles-Long Beach MSA. 

The ST tensor co-clustering method provides (i) a set of migration systems, (ii) a ranking of the core migration systems, and (iii) a temporal profile of the intensity of each migration system. At its crudest the ST tensor co-clustering method provides indicators of the associations of each US county with the six migration systems (in the way as described in Figure 1). The number of systems (i.e., $F$ in the model) is picked by observing the residuals between the low rank representation and the complete tensor data---see more discussions in the SI appendix. The low-rank decomposition aims to find the following representation of the spatial temporal data ${\cal X}$:
\begin{align*}
\mathcal{X} \approx \sum_{f=1}^F {\bf a}_f\circ {\bf b}_f\circ{\bf c}_f,~ \mathbf{A}\geq \bm 0, \mathbf{B}\geq \bm 0, \mathbf{C} \geq {\bm 0}, 
\end{align*}
where ${\bf A}=[{\bf a}_1,\ldots,{\bf a}_F]$ and ${\bf B}$ and ${\bf C}$ are defined in an identical way. 
The nonnegativity constraints are added to the factor matrices to reflect their physical meaning (i.e., the level of involvement in different migration systems for ${\bf A}$ and ${\bf B}$ and the activity intensity for ${\bf C}$). The columns of ${\bf A}, {\bf B}$ and ${\bf C}$ are normalized to have unit Euclidean norms; see details in the SI appendix.

\begin{figure}[H]
\centering
\includegraphics[width=.85\linewidth]{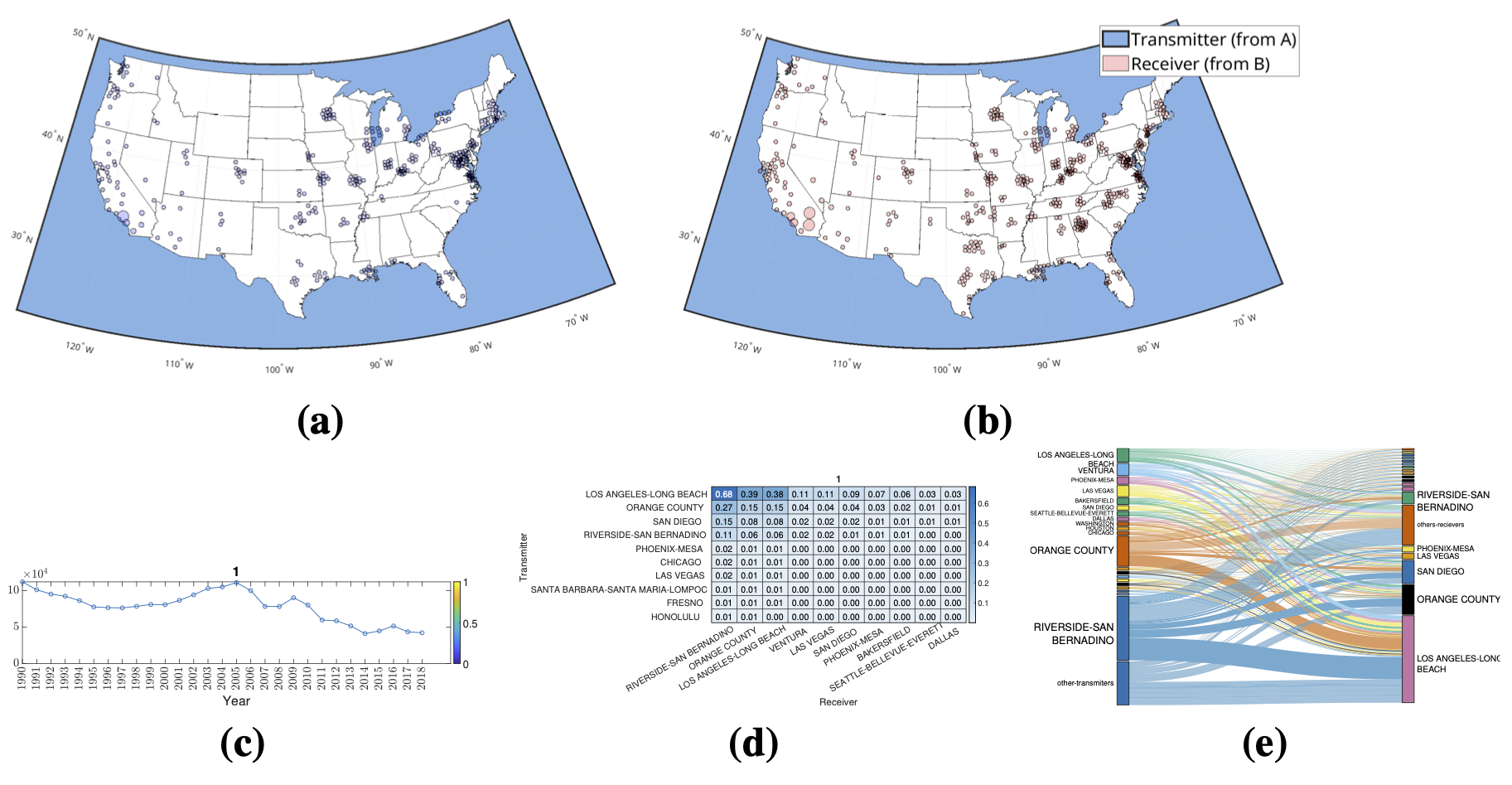}
\caption{Spatial and temporal plots of US Metro origin and destination migration systems. \textbf{(a)} Significant origin MSAs; \textbf{(b)} Significant destination MSAs; \textbf{(c)} temporal (intensity) profile of the migration system; \textbf{(d)} matrix representation of the migration system with inclusion with origin on the rows and destination on the columns; and \textbf{(e)} Sankey diagram of this migration system. 
}\label{fig:MSA_COM1}
\end{figure}

We focus on the top five to ten counties in each migration system as the probability of a county being in the system quickly approaches approximately zero in almost all cases below this threshold. See Figure~\ref{fig:MSA_COM1} where the core migration system is Los Angeles-Long Beach MSA, Orange County MSA, San Diego MSA and Riverside-San Bernardino MSA all in California with secondary set of MSAs in Arizona (Phoenix-Mesa), Illinois (Chicago), and Nevada (Las Vegas). Looking at Figure~\ref{fig:MSA_COM1}:\textbf{(c)} time-series plot we see that this method pulls out (in a fully data-driven way) the same qualitative story as \cite{frey2009great} which found that California lost most migrants to Arizona and Nevada in 2004-2005 and pre-housing collapse in 2010 a gain in migrants to Riverside-San Bernadino MSA from 2007-2009. We find that there is bump up in migration intensity as US housing recovers starting in 2013/2014 (see \cite{schuetz2021housing}) and decline as the housing market begins to heat up in 2016. We can observe the effect of this migration system in a Sankey diagram (Figure~\ref{fig:MSA_COM1}:\textbf{(e)}).

Similar to Frey \cite{frey1996immigration} we can observe the internal migration decline in this migration system dominated by the Los Angeles-Long Beach MSA, and later we can see a rebound in the early 2000s followed by the most recent decline.

\subsection*{IRS Data - California}

\begin{figure}[H]
\centering
\includegraphics[width=.85\linewidth]{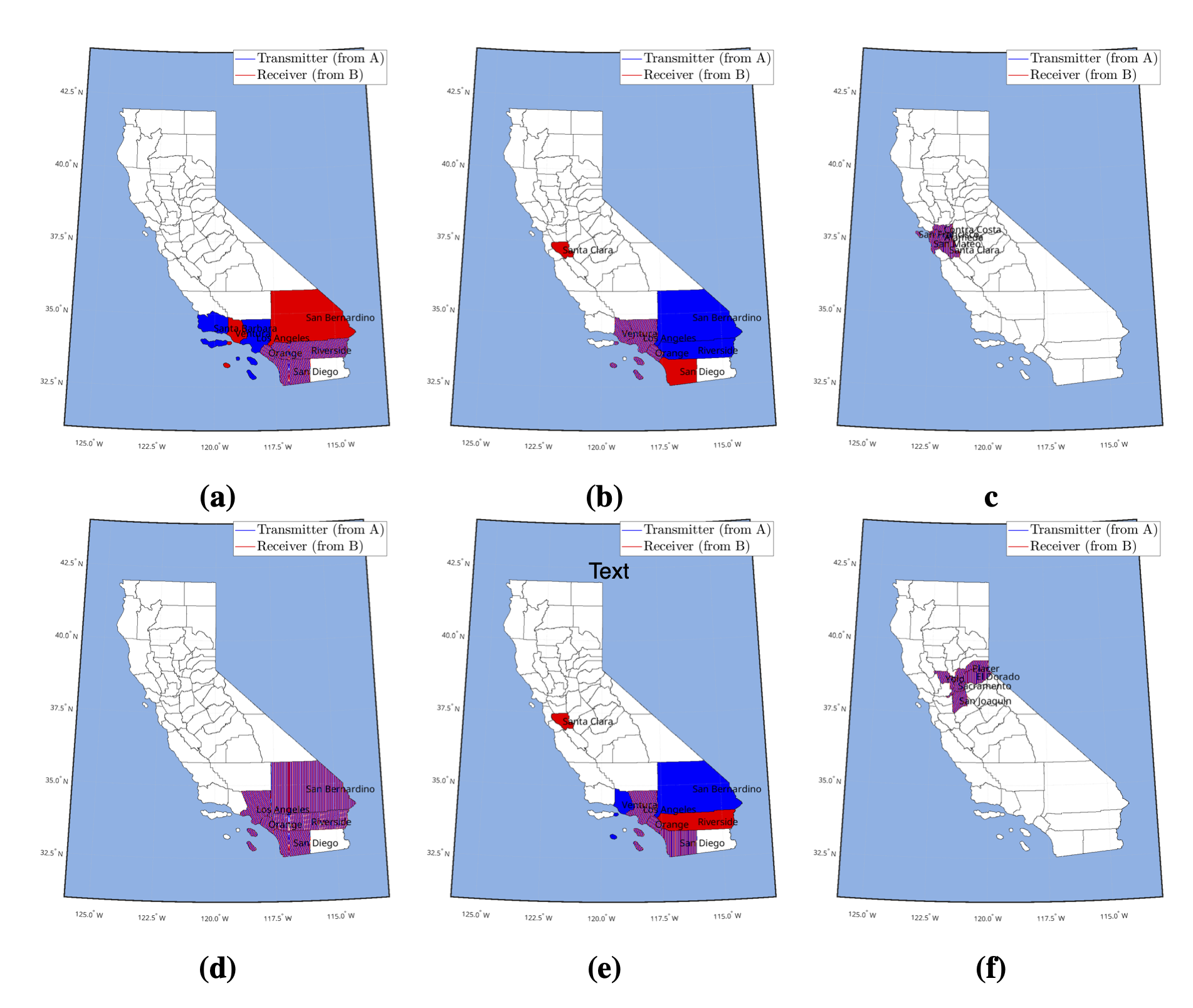}
\caption{ Major origin and destination migration systems (networks) for California Counties. The six tiles represent the six most important communities in relation to sending and receiving over 1990 to 2018. \textbf{(a)} Significant origin and destination counties in Community 1 \textbf{(b)} Significant origin and destination counties in Community 2; \textbf{(c)} Significant origin and destination counties in Community 3; \textbf{(d)} Significant origin and destination counties in Community 4;  \textbf{(e)} Significant origin and destination counties in Community 5; and \textbf{(f)} Significant origin and destination counties in Community 6.
}\label{fig:California_COM}
\end{figure}

California has been one of the most studied states for domestic migration \cite{frey1995immigration}. Further, Frey \cite{frey1995immigration} and others have shown that California is a high migration state with intense internal and cross-state effects. Here, we zoom into California over the period 1990-2018 and look at county-to-county migration just within the state. We observe two core communities that are defined by Southern California and Northern California. This aligns with a colloquial notion of the North/South divide in California popular culture (see Figure~\ref{fig:California_COM}). 
A basic question in the migration systems framework is which are the ``core migration systems'' and which are the most important. 
California is the largest state in the U.S. with approximately 40 Million residents and has two of the best known regions within a state: Northern California (centered at San Francisco/Bay Area) and Southern California (centered at Los Angeles). 

\begin{figure}[H]
\centering
\includegraphics[width=.85\linewidth]{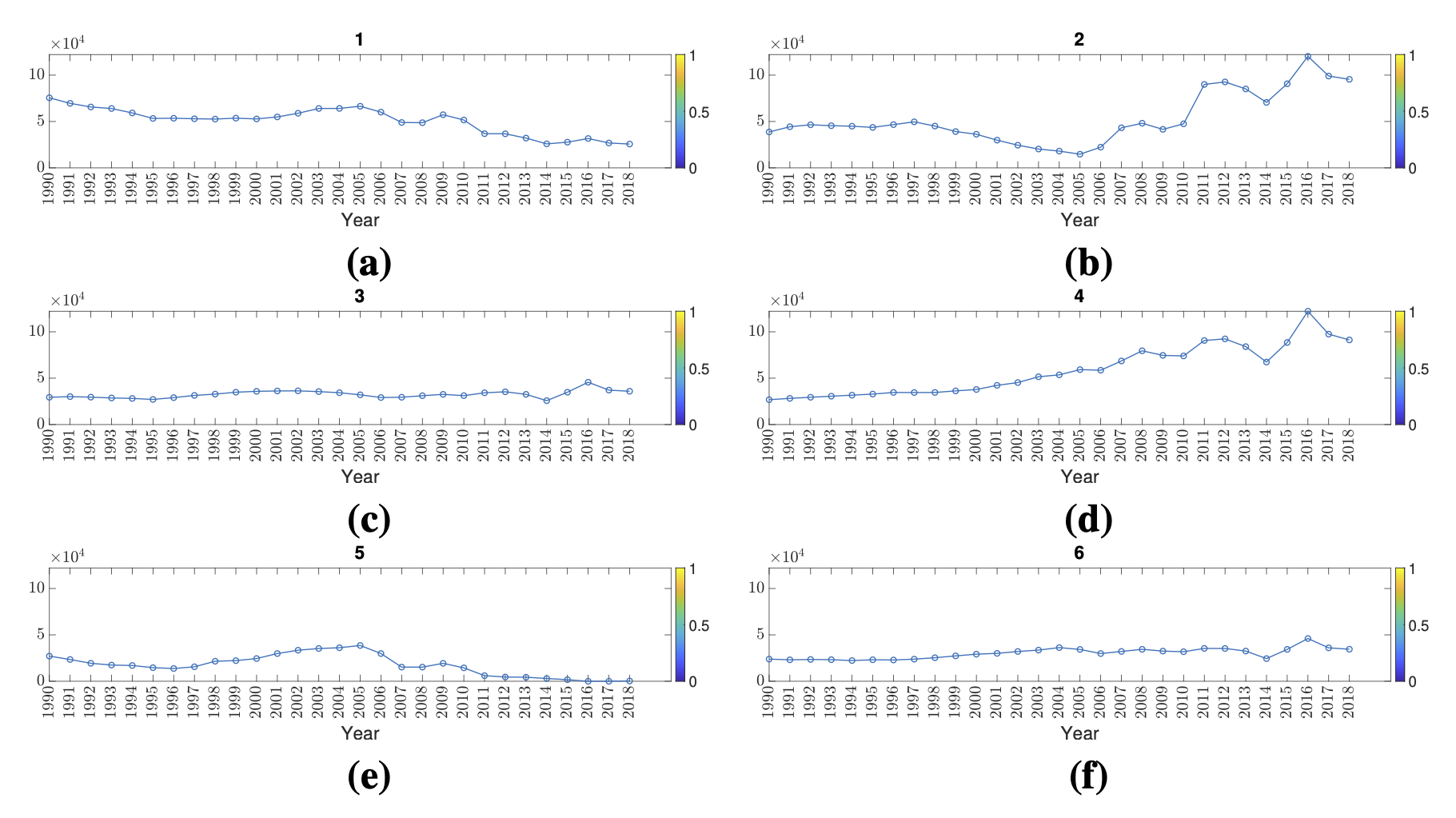}
\caption{For each of the six communities in Figure~\ref{fig:California_COM} there is a temporal profile for the ``intensity" of the migration system over time. This allows us to see how active this migration system is
at different time periods. \textbf{(a)} Temporal intensity profile for Community 1 \textbf{(b)} Temporal intensity profile for Community 2; \textbf{(c)} Temporal intensity profile for Community 3; \textbf{(d)} Temporal intensity profile for Community 4;  \textbf{(e)} Temporal intensity profile for  Community 5; and \textbf{(f)} Temporal intensity profile for  Community 6.
}\label{fig:California_TEMP}
\end{figure}

First, we focus on finding which counties form the major migration systems for movement within the state of California. We find two major systems (Southern California (communities 1-3 in Figure~\ref{fig:California_COM}: (a), (b) and (c)) and Northern California (communities 4-5 in Figure~\ref{fig:California_COM}: (d) and (e))) with Los Angeles as the link between the two systems. Next, we look at the temporal profiles (Figure~\ref{fig:California_TEMP}) of the migration systems. We discover that in community 3 (Figure~\ref{fig:California_TEMP}:(c)) the temporal intensity matches the crest of unemployment and subsequent decline in unemployment (see \cite{tan2021california}), which reinforces the idea of the importance of labor markets on internal migration. Focusing on the temporal profiles of these migration systems, we see that the core migration system (community 1; Figure~\ref{fig:California_COM}) shows the general trend known as the ``Great American Migration Slowdown'' (coined by Frey \cite[p.1]{frey2009great})). It is generally established that there has been decline in internal migration since about the 1970s, with the slowdown picking up in the 1990s \cite{molloy2011internal}. From the figure we also pick up the decline in unemployment from around 2010 to 2018 (see \cite{tan2021california}).

\begin{figure}[H]
\centering
\includegraphics[width=.85\linewidth]{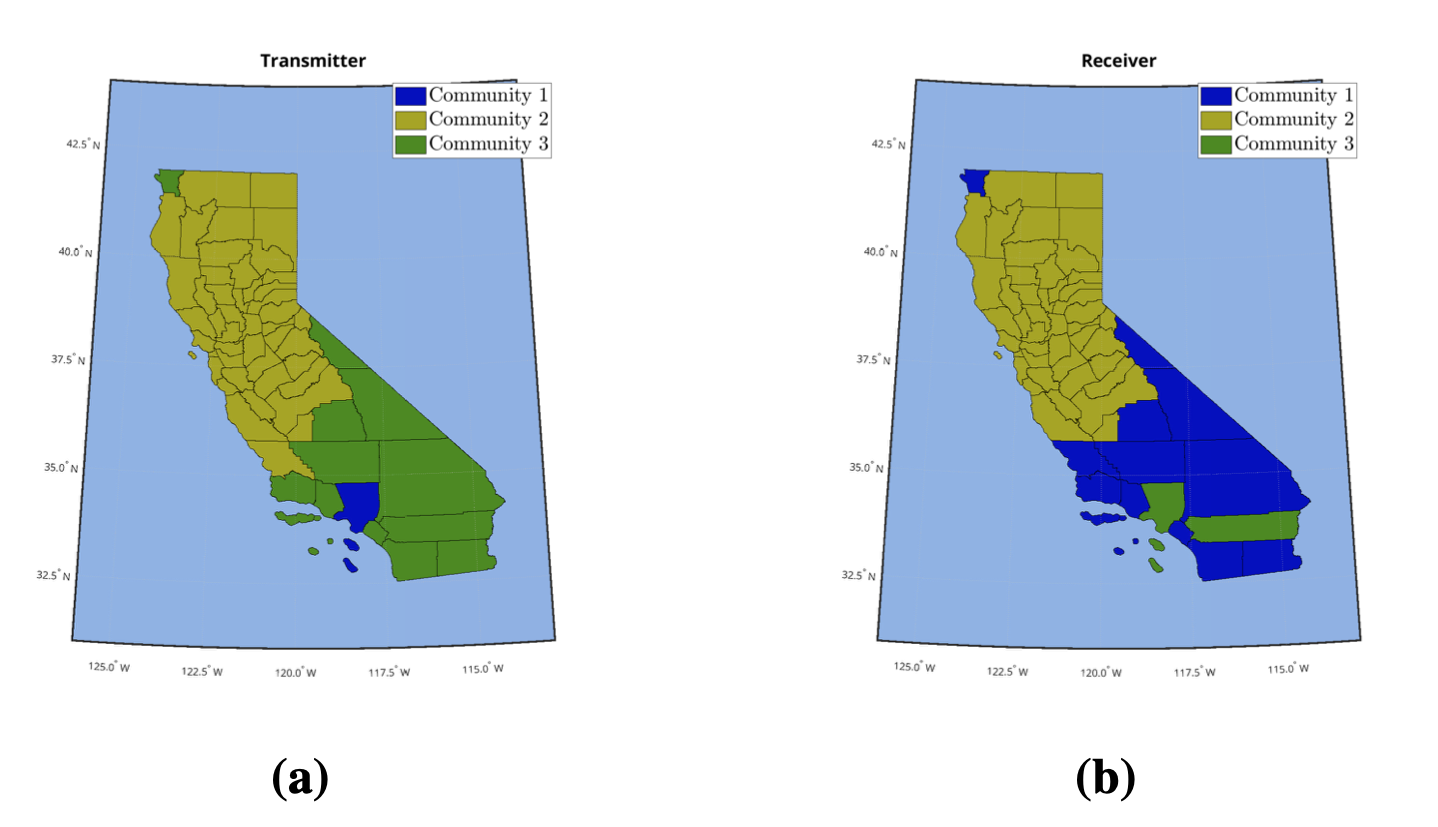}
\caption{These two plots show a fully classified California for origin/destination migration systems under a three system model. \textbf{(a)} Sender (origin) migration systems (3 Community solution); and \textbf{(b)} receiver (destination) migration systems (3 Community solution).
}\label{fig:California_3COM}
\end{figure}

Given these two systems we might be interested in classifying the whole state as three community systems using our method combined with a clustering algorithm (in this case k-nearest neighbors; \cite{fukunaga1975branch}). In Figure~\ref{fig:California_3COM} we can see the Northern California versus Southern California split with Los Angeles being the core origin/destination system for Southern California.  

\subsection*{IRS Data - Hurricane Katrina: Louisiana}

\begin{figure}[H]
\centering
\includegraphics[width=.85\linewidth]{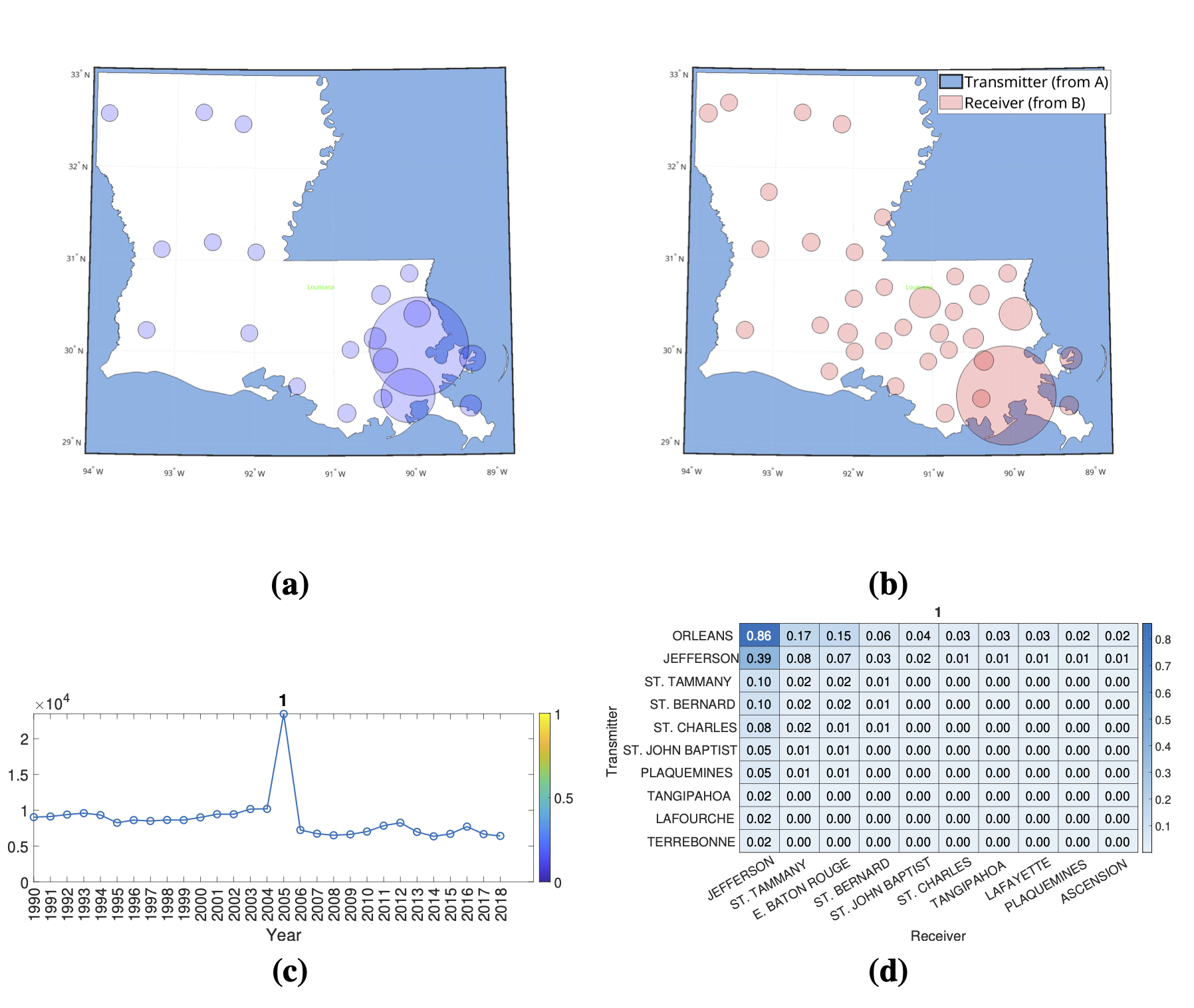}
\caption{Spatial and temporal plots of the first migration system (community 1) obtained using the proposed ST tensor method applied to Louisiana. Illustrates a large temporal shift at the point of Hurricane Katrina in 2005. \textbf{(a)} Significant sending (origin) counties in the migration system (community 1); \textbf{(b)} Significant receiving (destination) counties in the migration system (community 1); \textbf{(c)} temporal profile of the migration system, 1990-2018 (community 1); and \textbf{(d)} node-to-node intensity for the migration system (community 1).
}\label{fig:NOLA_COM1}
\end{figure}

\begin{figure}[H]
\centering
\includegraphics[width=.85\linewidth]{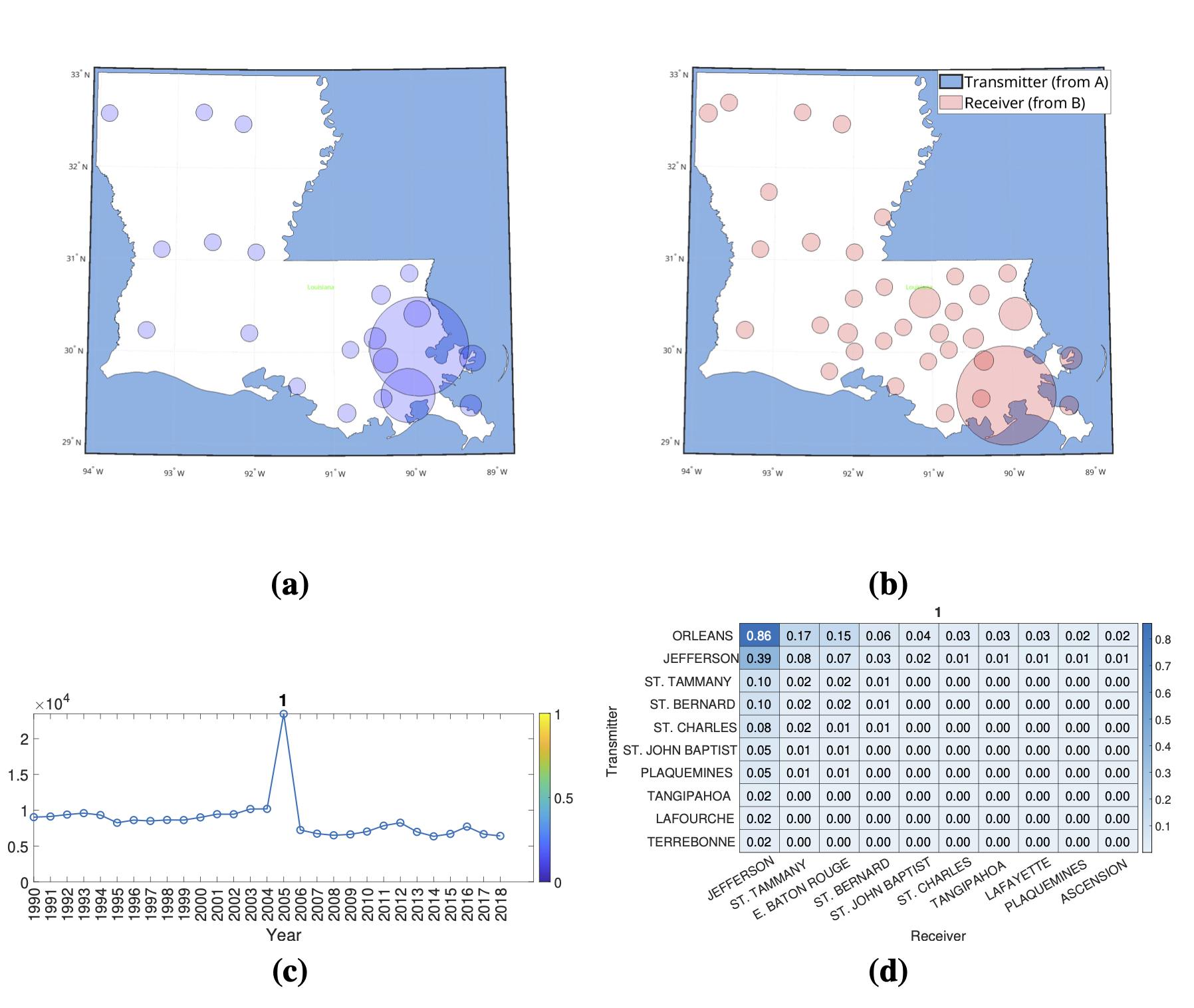}
\caption{Spatial and temporal plots of the first migration system (community 2) obtained using the proposed ST tensor method applied to Louisiana. Illustrates a large temporal shift at the point of recovery from Hurricane Katrina in 2005. \textbf{(a)} Significant sending (origin) counties in the migration system (community 2); \textbf{(b)} Significant receiving (destination) counties in the migration system (community 2); \textbf{(c)} temporal profile of the migration system, 1990-2018 (community 2); and \textbf{(d)} node-to-node intensity for the migration system (community 2).
}\label{fig:NOLA_COM2}
\end{figure}

One particularly compelling aspect of the ST tensor co-clustering method is the ability to detect the activity intensity changes in the migration system in response to external shocks. In 2005, Hurricane Katrina's effect on the City of New Orleans provided an extreme example of how severe weather events can change the demographics of a major city \cite{fussell2014recovery,dewaard2016population}. Here we are focused on the migration system between New Orleans Parish and all other counties within the state of Louisiana and a node representing all the combined counties outside the state. Pre-disaster is defined as before 2004 and recovery as 2007-2009 \cite{dewaard2016population}. In Figure~\ref{fig:NOLA_COM1} (community 1), we can clearly see the migration out of New Orleans Parish. In Figure~\ref{fig:NOLA_COM2} (community 2), we can see the recovery cresting in 2009. This analysis demonstrates a particularly important feature of the ST tensor co-clustering in that it allows us to see which  migration systems are activated due to exogenous events (e.g. natural disasters) and which ones are activated for recovery from exogenous events. In this case we see clearly that the system that is engaged after the disaster (community 1) is different than the one that is engaged for recovery (community 2; though they do share some counties in common). These changes to the migration system have been further analyzed in the SI Appendix. The key finding here is that we can see exactly which migration systems are activated for the displacement event (Hurricane Katrina) and which systems are activated for return migration (recovery). Further, we can see that while there is overlap in the counties, it is not the same communities that are involved in the recovery -- suggesting that that some of the recovery is driven by new migration to the area.

\paragraph{Comparison with alternative methods} A typical strategy for dynamic graph clustering is to apply a classic `static' community detection technique designed for graphs without the temporal dimension (such as the walktrap method \cite{pons2005computing}) to the graphs collected at different time-points separately and look at how the resulting system changes. Here we use a walktrap algorithm on the Louisiana domestic migration network split into pre- and post- Hurricane Katrina. We then compare the ST co-clustering method with the walktrap method (see details in Materials and Methods). The walktrap method is a random walk based community detection algorithm. It provides hard nonoverlapping clustering of the counties based on their migration patterns. However, the method does not tell which migration system exhibits higher activity levels as revealed in our method; see (Figure~\ref{fig:Pre-Post-Katrina-NewOrleans}). To provide a better comparison with the ST co-clustering method we observe the cluster (i.e., a migration system) that contains New Orleans, which is the prominent city and the one hit heavily by Katrina (Figure~\ref{fig:Pre-Post-Katrina-NewOrleans}).  In Figure~\ref{fig:sub:our_cpd} we look at the top 5 origin (sender) and destination (receiver) counties found by the ST tensor co-clustering method. Note that walktrap does not offer such sender/receiver information. Next, we observe differences in cluster patterns with communities 1 and 2 producing the closest to the walktrap solution (see Figure~\ref{fig:Pre-Post-Katrina-NewOrleans}). According to walktrap method, the primary system in New Orleans shrinks by half between pre- and post-Hurricane Katrina representing the changes in the system. However, in the ST co-clustering method we see the local cluster around New Orleans with only East Baton Rouge Parish being in the system suggesting that the evacuation was much closer to the disaster center than the walktrap method would suggest.

\begin{figure}[H]
    \centering
\includegraphics[width=.85\linewidth]{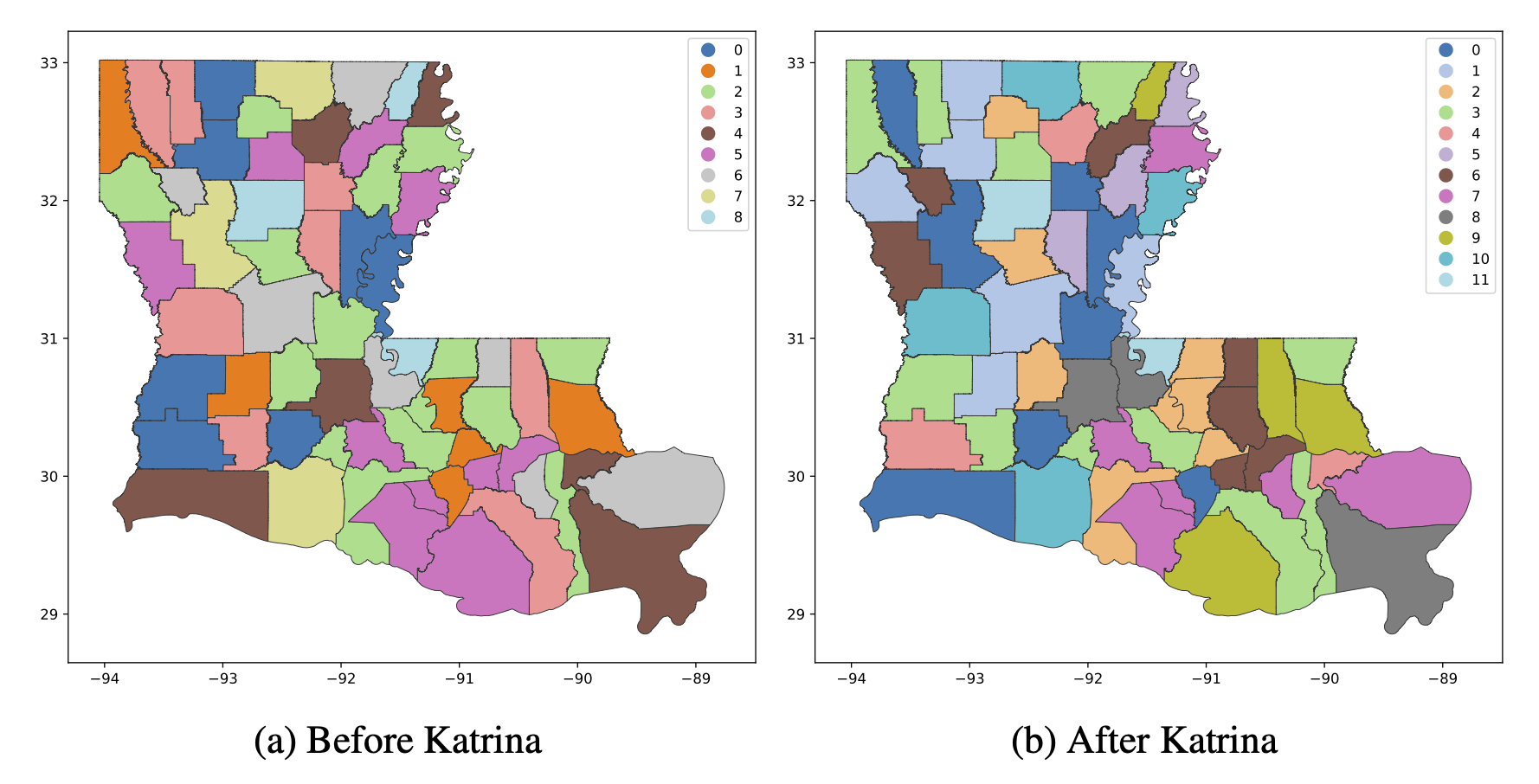}
    \caption{Complete partitioning pre and post Hurricane Katrina in 2005 based on the walktrap community detection solution. }
    \label{fig:Pre-Post-Katrina-RandonWalk}
\end{figure}

\begin{figure}[H]
    \centering
 \includegraphics[width=.85\linewidth]{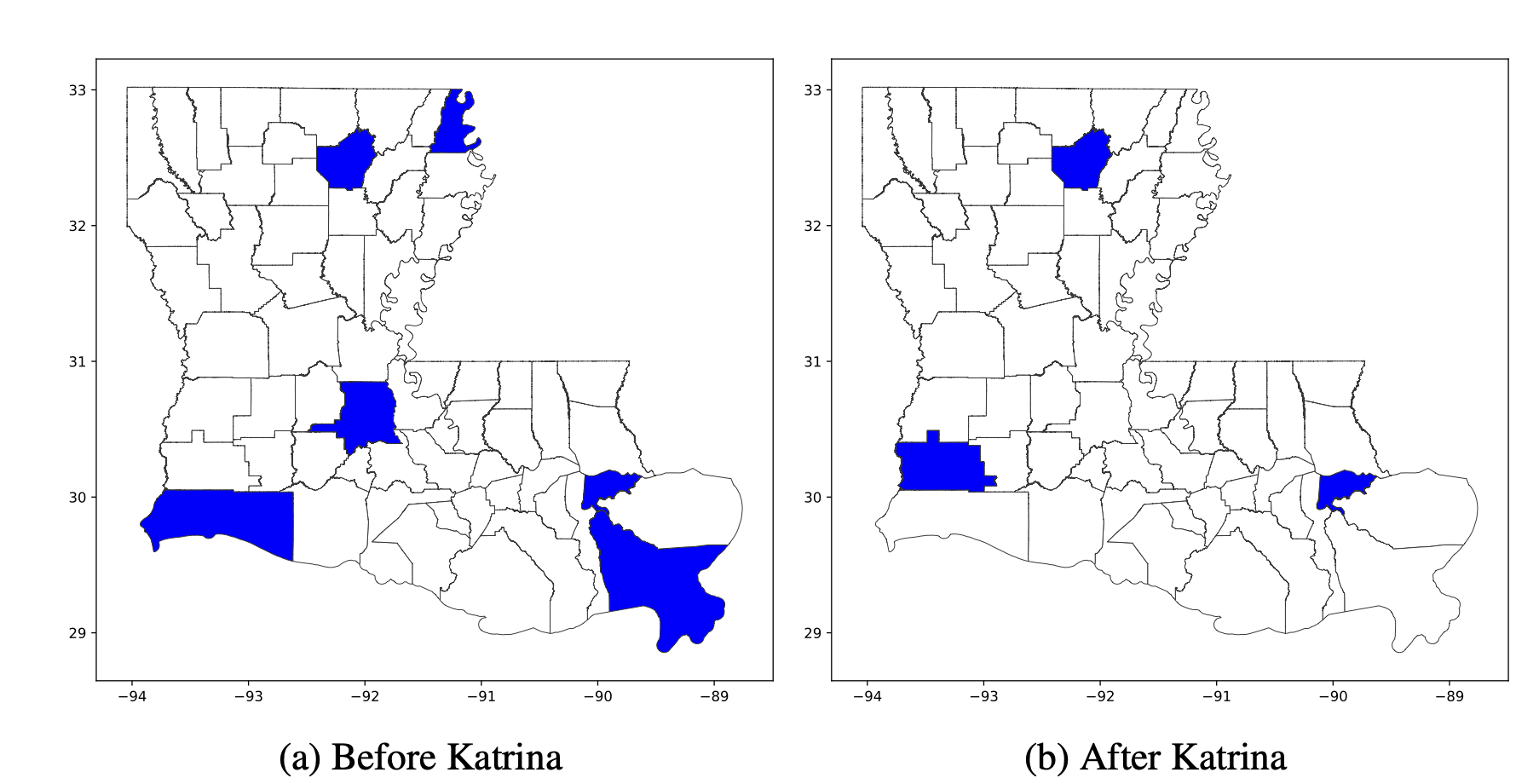}    
 \caption{New Orleans Parish community as established pre and post Hurricane Katrina in 2005 based on the walktrap community detection solution. }
        \label{fig:Pre-Post-Katrina-NewOrleans}
\end{figure}


\begin{figure}[H]
    \centering
\includegraphics[width=.85\linewidth]{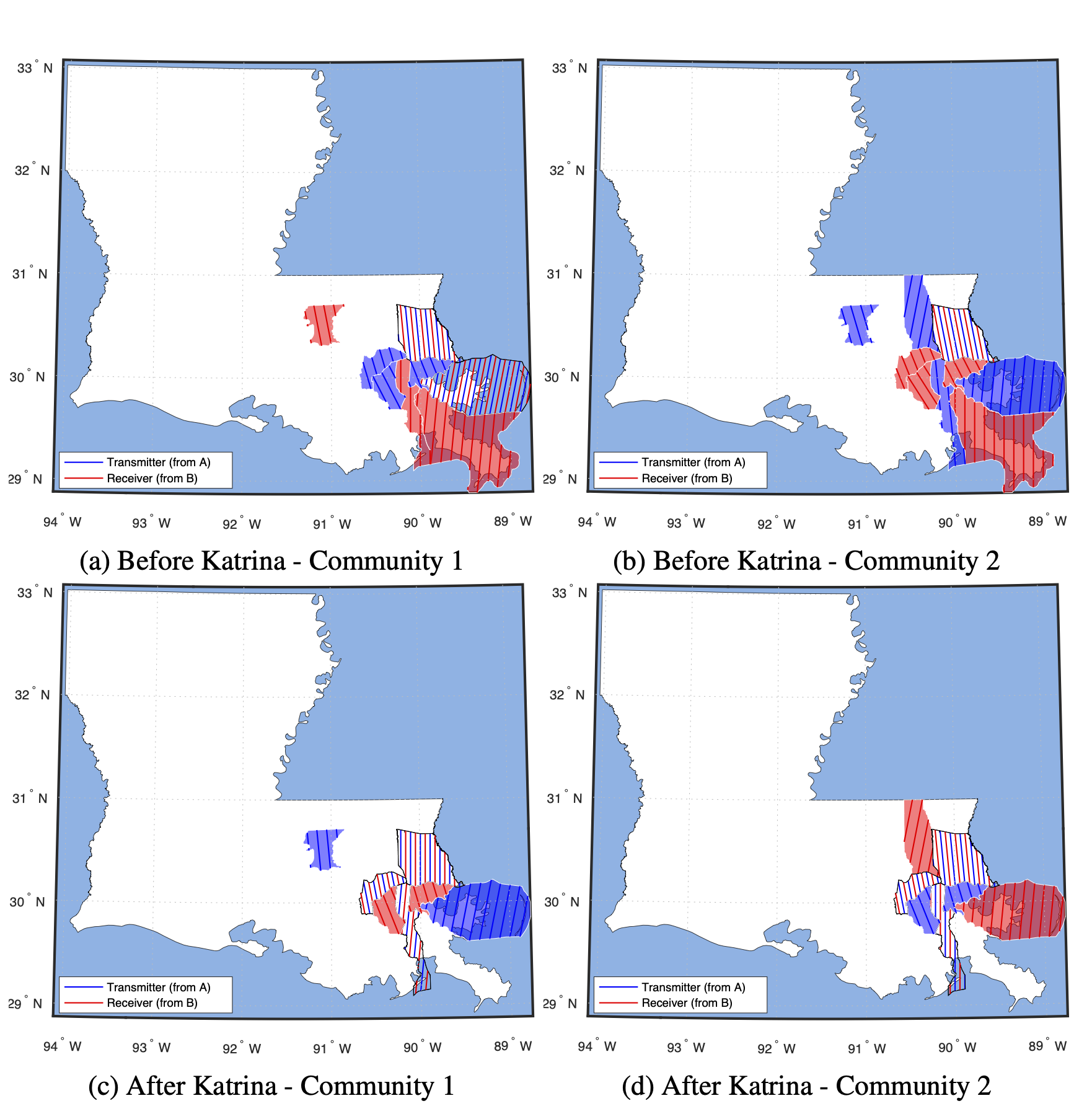}
    \caption{The New Orleans community output by the ST tensor co-clustering method. Communities are plotted with their top 5 transmitters and top 5 receivers. It happens that New Orleans appears in top 5 transmitters/receivers in community 1 and 2 (which are top 2 significant communities).}
    \label{fig:sub:our_cpd}
\end{figure}

\subsection*{International Migration Data - Global Migration Systems}

The identification of migration systems in the international context continues to be an open problem in the field. A number of works have posited that there should exist international migration systems \cite{bakewell2014relaunching,kritz1992international,mabogunje1970systems,massey1999worlds} and have provided a set of ``general principles" of such systems
rather than analytic approaches \cite{abel2021form}. Recent work by Abel et al. \cite{abel2021form} has applied static community detection methods to the international migration data provided by Azose and Raftery \cite{azose2019estimation} and updated by Abel et al. \cite{abel2021form} to demonstrate the change in migration systems over time. In \cite{abel2021form}, the migration systems were found in a year by year manner, by repeatedly applying the community detection method to each year's data.

Here, we apply the ST tensor co-clustering method to the same data. This results in some important differences in output and understanding of the migration systems. First, ST tensor co-clustering produces a set of migration systems which exist over the entire period (though intensity of their importance varies over time) and thus represent a cleaner set of migration systems than those in Abel et al. \cite{abel2021form}. By fitting community detection methods year by year, Abel et al \cite{abel2021form} cannot guarantee the persistence of a given system. 
This means that the method does not ensure discovery of clusters (networks) of countries that keep similar spatial interactions with varying activity intensity over time.
Thus, the ST co-clustering model produces a better representation of the migration system, especially under what Kritz and Zlotnik \cite{kritz1992international} described as ``network[s] consisting of sets of concept of dynamic stability'' (see also \cite{kritz1992international}). 

Here, we explore six major communities consisting of the top 10 origin and destination locations over 1990 to 2015 (Fig.~\ref{fig:world} and \ref{fig:world2}). This set is chosen based on least squares fit of the data (see the appendix for details). The first migration system (Fig.~\ref{fig:world}; community 1) is dominated by the relationship between Mexico (origin) and the United States (destination) and is characterized by countries sending migrants to the United States. The second migration system  (Fig.~\ref{fig:world}; community 2) is characterized by Eastern European migration with Russia dominating both the origin and destination of the the system. The third migration system (Fig.~\ref{fig:world}; community 3) is characterized by India and Bangladesh and primarily other counties in Southern and South Eastern Asia. The fourth migration system (Fig.~\ref{fig:world}; community 4) is characterized by the United States, China and India as a largest origin countries with destination countries Mexico, South America, Asia and Western Europe and Russia as the primary set. The fifth migration system (Fig.~\ref{fig:world}; community 5) is dominated by the Middle East with Syria being the largest origin country. Last, the sixth migration system (Fig.~\ref{fig:world}; community 6) is also in the Middle East and is composed of Iran, Pakistan and Afghanistan. 

We have visualized these communities with world maps in Fig~\ref{fig:world2}. In Fig.~\ref{fig:world3}, we provide examples of the temporal profile and the spatial interaction matrix of the top 10 countries in migration system 1 to 6. 
The matrix is produced by instantiating $\widetilde{\bf a}_f(i)\widetilde{\bf b}_f(j)$ as its $(i,j)$th element, where $\widetilde{\bf a}_f(i)\in\mathbb{R}^{10}$ and $\widetilde{\bf b}_f(i)\in\mathbb{R}^{10}$ are the sub-vectors of ${\bf a}_f$ and ${\bf b}_f$ holding the top-10 strongest elements, respectively.
The temporal profile allow us to see major events such as the end of Syrian occupation of Lebanon in 2005 (Community 4; Fig.~\ref{fig:world3}). Altogether, the results are similar to those in the work of Abel et al. \cite{abel2021form} with major migration systems centering around the United States with distinct Europe and Eastern European systems, Middle East and Southern Asia systems. Further, we find evidence of change in importance of the United States dominated migration system from migration system 1 (Community 1; Fig.~\ref{fig:world3}) to migration system 3 (Community 3; Fig.~\ref{fig:world3}) in 2005-2010, which is similar to what Abel et al. \cite{abel2021form} found. However, we can see that the process of this change started at the beginning of the period (Community 3; Fig.~\ref{fig:world3}) with the peak change occurring in 2005-2010. Further, because we have a stable set of countries in our system we can see specifically how and when one system versus another becomes dominant, from the temporal profile change of two US dominated systems. Another key differences include our ability to spot major changes (e.g. end of Syrian occupation of Lebanon in 2005)---which does not show up in the work of Abel et al. \cite{abel2021form}.

\begin{figure}[H]
\vspace*{-1in}
\centering
\includegraphics[width=.8\linewidth]{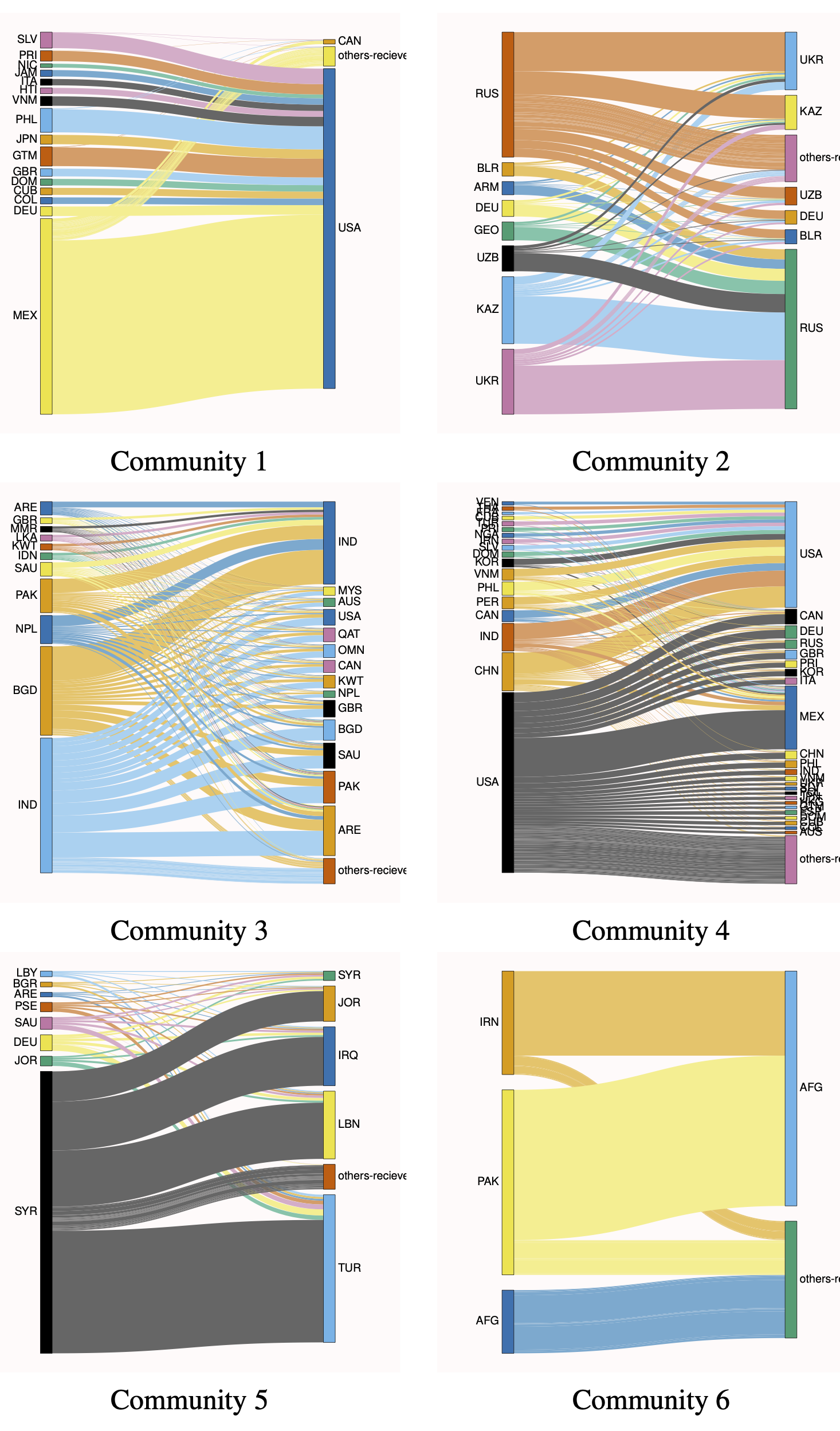}
\caption{Top six migration systems for international migration for the period 1990-2015. On the left is the main origin country in the system and on the right is the main destination country for the system. The size of the path between countries represents the scale of the sending or receiving migration pipeline between the two countries.
}\label{fig:world}
\end{figure}

\begin{figure}[H]
\vspace*{-1in}
\centering
\includegraphics[width=.85\linewidth]{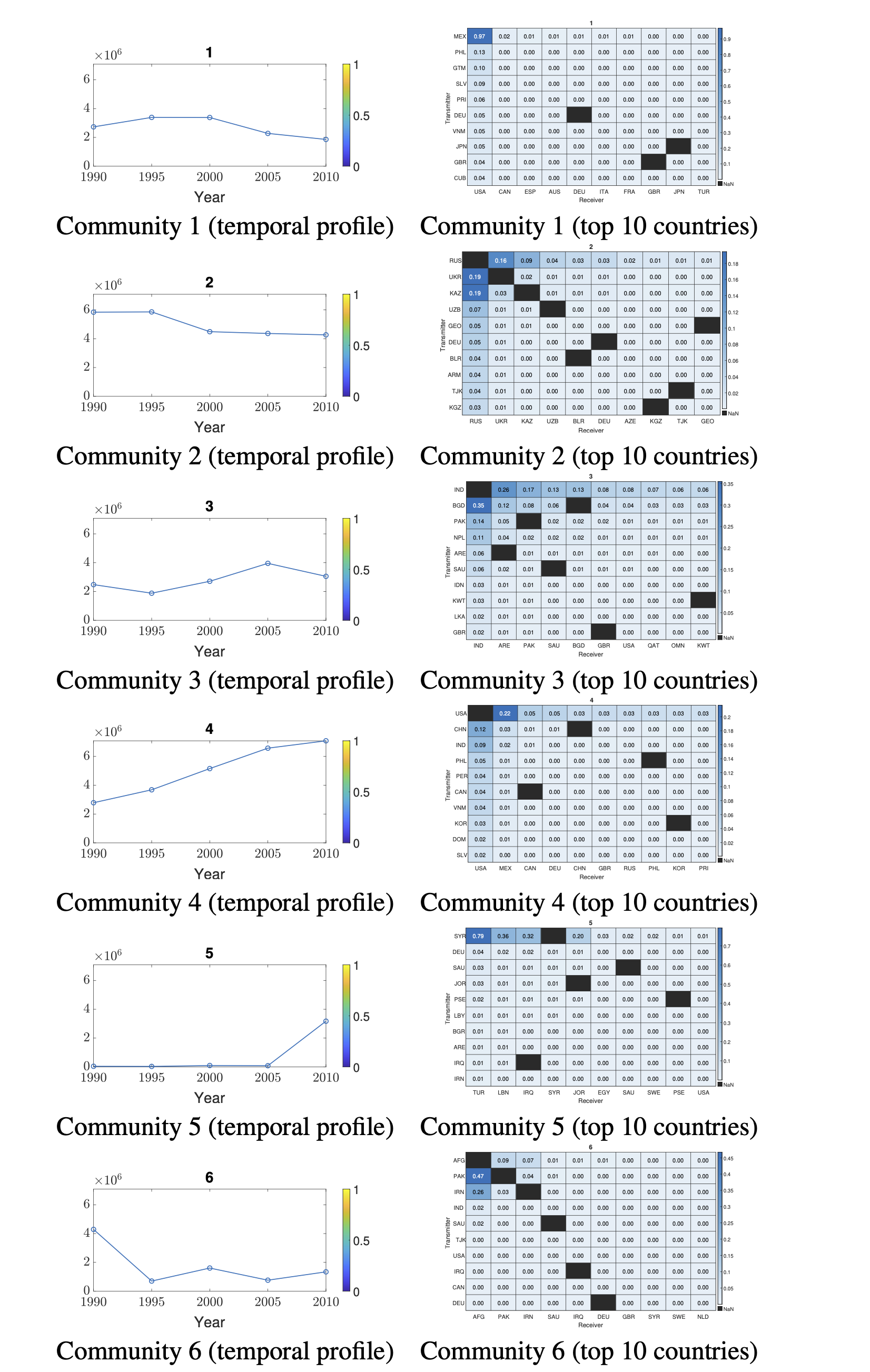}
\caption{The temporal profile (left) and top 10 countries (right) for the six migration systems estimated from the international migration data from 1990-2015. All years on the label are coded from the first year in the five year interval, i.e. 1990-1995 is labelled 1990 and 2010-2015 is labelled as 2010.}\label{fig:world3}
\end{figure}

\begin{figure}[H]
    \centering
\includegraphics[width=.85\linewidth]{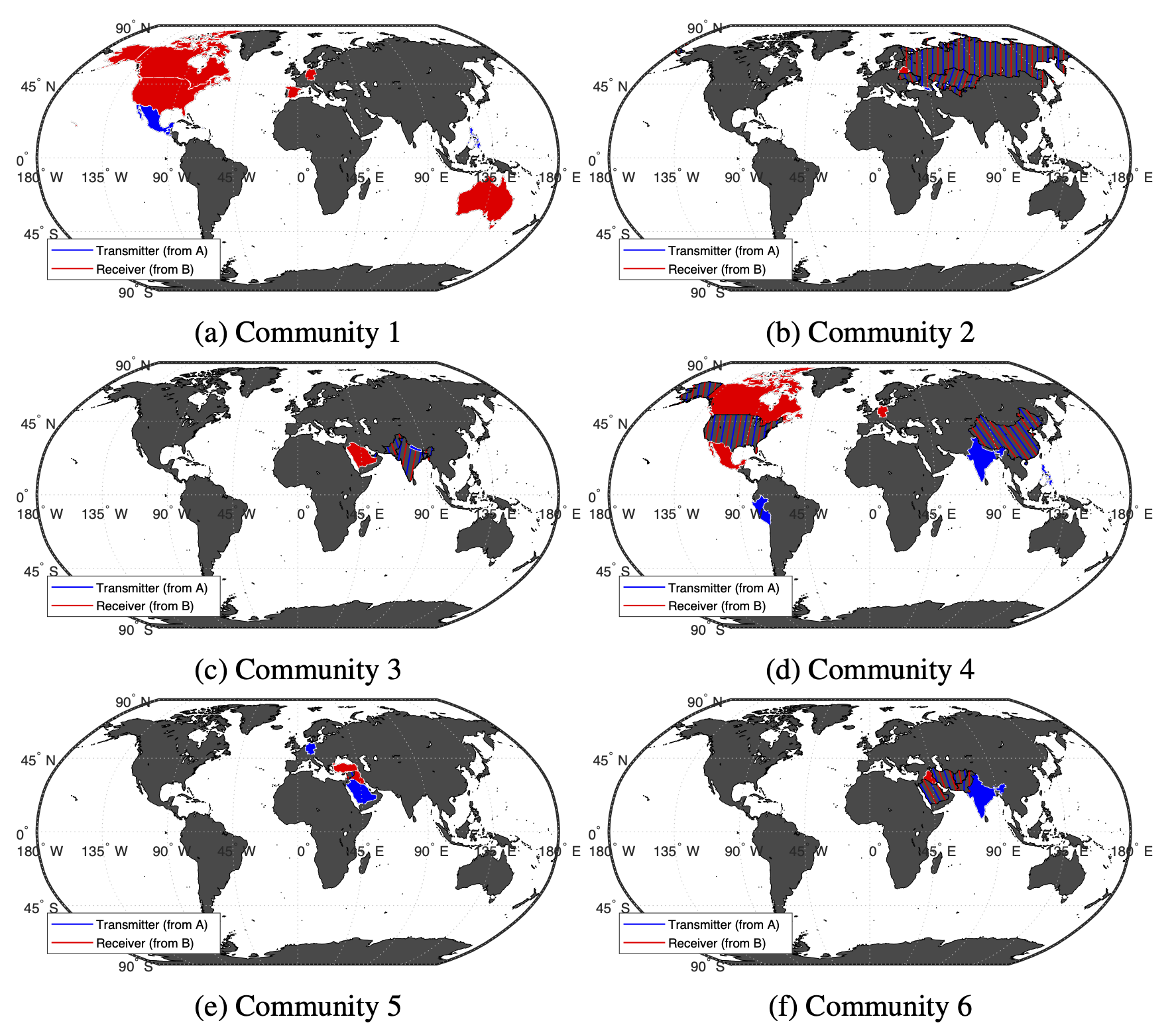}
\caption{Top six migration systems for world migration for top 10 origin and destination countries from the ST tensor co-clustering method. Red is a significant destination country and blue is a significant origin country in the migration system. Striped indicates the country is both a significant origin and destination country in the migration system. 
}
\label{fig:world2}
\end{figure}

\begin{figure}[H]
    \centering
  \includegraphics[width=.85\linewidth]{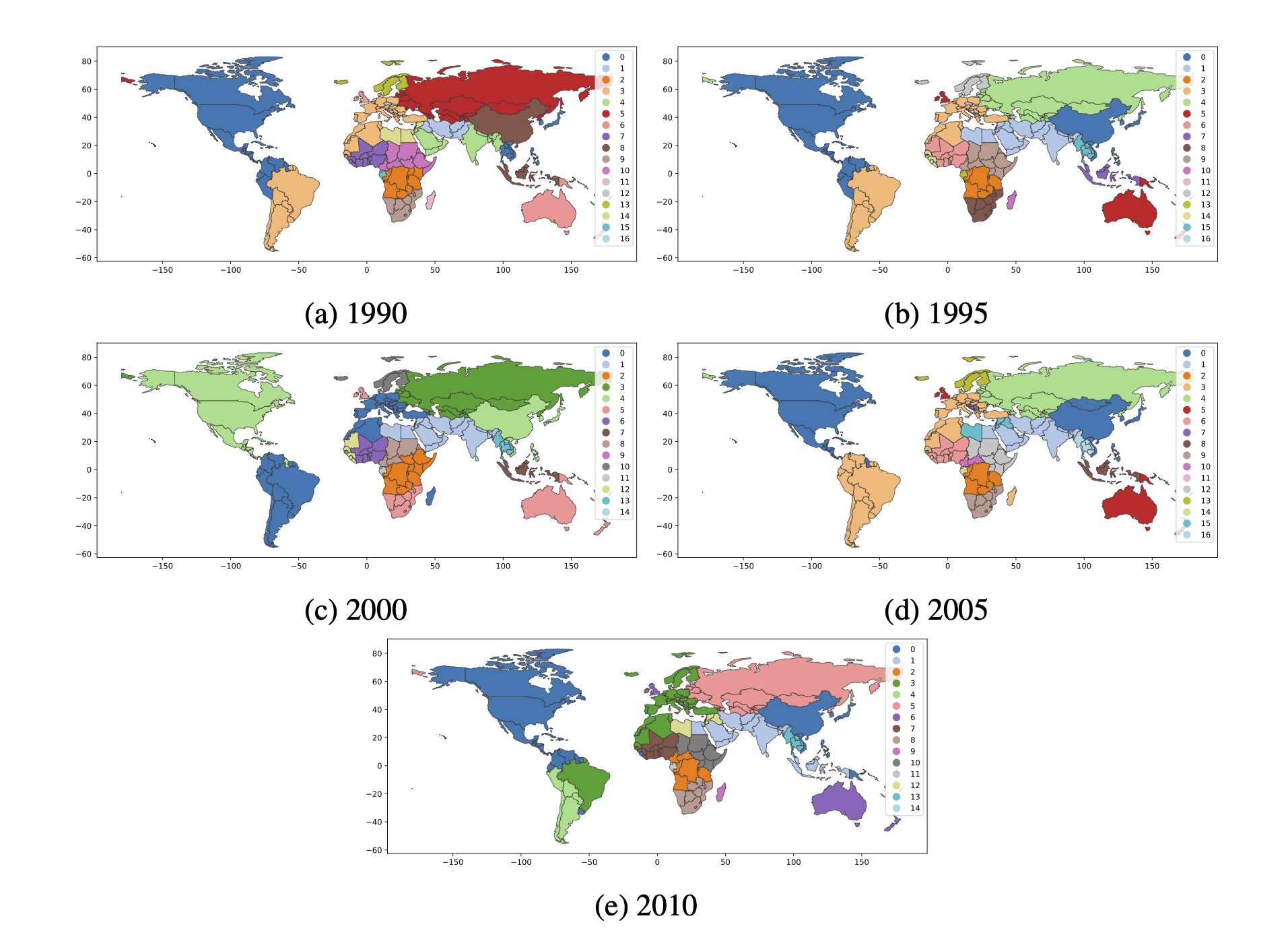}
    \caption{Migration systems discovered by method infoMap \cite{rosvall2008maps} for international migration for the period 1990-2015.}
    \label{fig:rosvall2008maps_full}
\end{figure}

\begin{figure}[H]
    \centering
  \includegraphics[width=.85\linewidth]{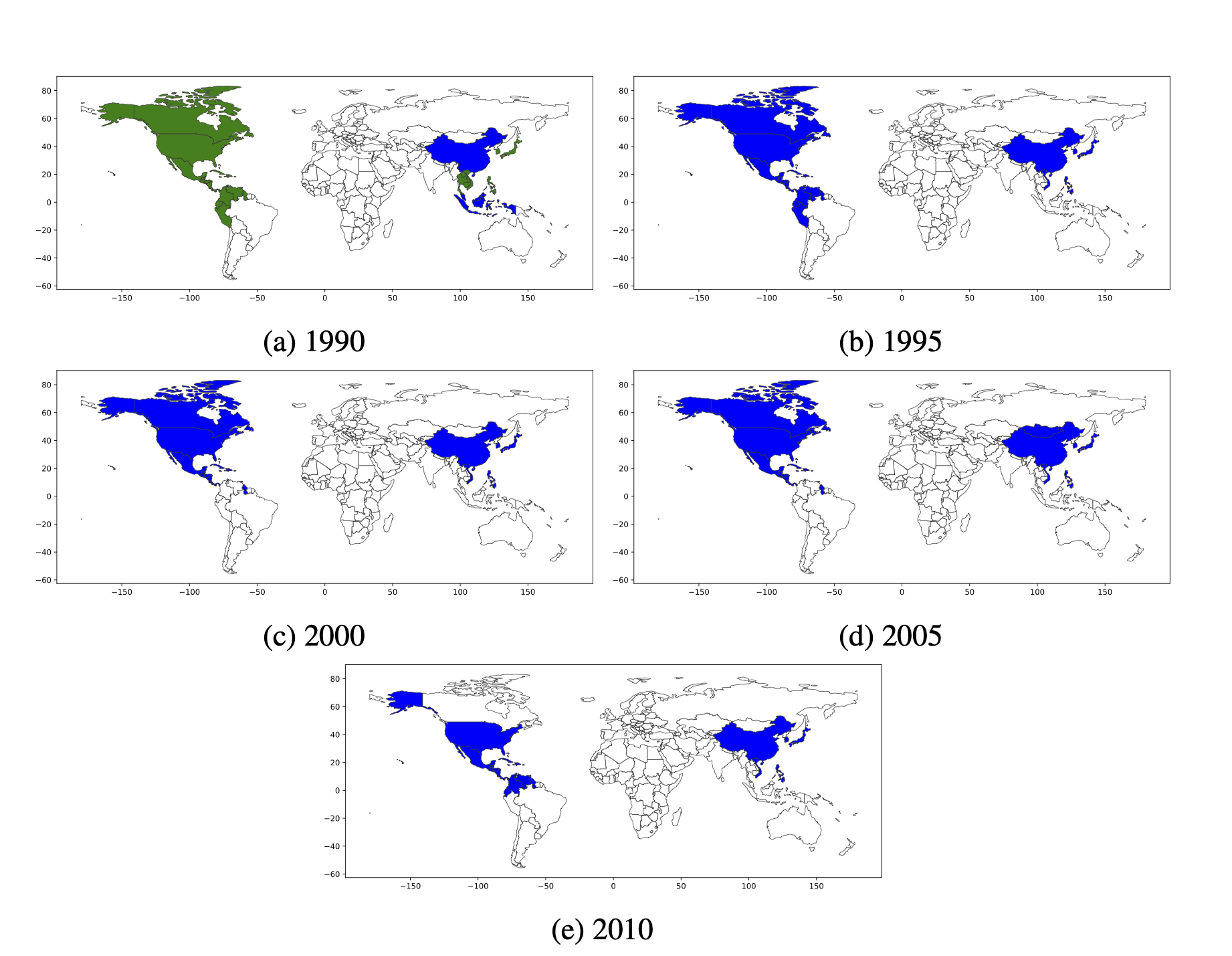}
    \caption{A subset of communities selected by requiring US or China to be in the cluster from Figure~\ref{fig:rosvall2008maps_full}. Green and Blue represent distinct clusters. In 1990 the US and and China are in distinct clusters and from 1995 to 2010 they are contained in the same cluster.
    }
    \label{fig:rosvall2008maps_small}
\end{figure}

\paragraph{Comparison with prior works} 
The work in Abel et al \cite{abel2021form} employed infoMap \cite{rosvall2008maps} to demonstrate the efficacy of finding migration systems using static network clustering techniques. 
Both methods generate a distinct North American cluster, European and Asian cluster; however the method of in \cite{abel2021form} does not provide distinct origin and destination clusters or highlight which communities are more significant over a given time period. We have recreated these clusters in Figure~\ref{fig:rosvall2008maps_full} and isolated the communities that include either the United States or China (Figure~\ref{fig:rosvall2008maps_small}). Again, one distinct difference is that the classic methods provide a full partitioning and focus on how these communities evolve over time, whereas our method provides a quantitative measure and ordering of how ``important'' a community is, and how stable the community is over time. Both methods detect the US-Canada-Mexico cluster, but our method also detects separately the China-US origin-destination cluster, where we can distinguish between whether the origin or destination is driving the relationship (see Figure~\ref{fig:rosvall2008maps_small} in comparison with Figure~\ref{fig:world2}). In the Appendix (Figure~S31), we have included a 3- and 6-cluster forced-partitioning solution from our model which captures the stable trends over time and separate origin and destination compared to the 5 infoMap solutions (Figure~\ref{fig:rosvall2008maps_full}). Altogether, the ST Co-Clustering method provides a distinct and useful solution differing from the current focus on change. Instead, ST Co-Clustering aims to uncover migration systems that are stable over time, as hypothesized in the literature, and further allows the ability to rank / prioritize these systems, as well as distinguish between origin and destination clusters.

\section*{Discussion and Summary}

Our findings have five important implications: (i) empirically derived migration systems can be established as stable over time and space (domestically and international migration), (ii) we can correctly derive expected migration systems across the US (e.g., Northern and Southern California) and international contexts, (iii) we can detect exogenous shocks to the migration system (e.g., Hurricane Katrina or the end of Syrian occupation of Lebanon) and (iv) we can establish changes in migration systems over time (e.g. Syrian Refugee Crisis) and (v) we provide a novel approach to dynamic community detection that focuses on stable clusters over time rather than change in clusters over time. Beyond the purely descriptive, these quantitative data-driven methods have the potential to improve population forecasting \cite{raftery2012bayesian}, as Andris et al. \cite{andris2011predicting} demonstrated the usefulness of migration clusters in predicting future migration flows or potentially could be integrated in formal demographic models such as those used in Massey et al. \cite{massey1999dynamics}. In general this is a powerful method for finding spatio-temporal clusters in networks such as migration systems. This method has broad potential impact in the social sciences beyond migration systems in areas such areas partisan politics or alliance formation. 

\section*{Materials and Methods}

\paragraph{IRS Data}

The IRS migration data are created in the following manner: (1) taxpayer identification numbers (TINs) are used to match tax returns in consecutive years; (2) matched tax returns where migrant returns are defined as those that do not match the state or county of residence in consecutive years; (3) total counts of tax returns and tax exemptions -- effectively households and individuals respectively -- and the total adjusted gross income or AGI contained in the migrant and non-migrant returns are aggregated up to the state and county levels \cite{hauer2019irs,gross2005internal,pierce2015soi}. There are four major limitations \cite{hauer2019irs} of this data set discussed in the literature that include: (1) by definition, these data do not include those who do not file a tax return. This group is disproportionately elderly and/or poor \cite{hauer2019irs}. (2) The data is limited to aggregate counts of county (or state) data on three variables (i) total counts of migrant and non-migrant returns (i.e. households), (ii) exemptions (i.e. individuals), and (iii) AGI. (3) There is a methodological change between 1990-2011 data and 2012-2018 data \cite{dewaard2021user}. In the 2011-2012 tax year, the preparation of the data shifted from the US Census Bureau to the IRS who expanded the window for returns included in the estimates to go through December rather than September. (4) The last criticism is that for privacy reasons the county-to-county migration flows involving less than 10 households are obscured \cite{dewaard2021user}. This was increased to 20 in the 2011-2018 periods \cite{dewaard2021user}. 

Dewaard et al. \cite{dewaard2021user} describe the limitations of using the combination of the 1990-2018 period due to processing standards change from US Census Bureau to IRS workers. Nonetheless, because the ST tensor co-clustering method works in a low-rank approximation manner that is analogous to the principal component analysis (PCA) for matrix data, such measurement error-induced noise is not expected to cause visible issues. Further, to evaluate the robustness of this assumption we have done a series of tests (available in the SI Appendix) with no major red flags showing up. So while \cite{dewaard2021user} does caution against this practice we find that our procedure is robust to the issues discussed. 

Alternative migration data in the US lacks temporal and spatial resolution for such an analysis. For example there exists a one year migration estimates from the 2000 US Census long form (1 year of data) and five year interval estimates from the American Community Survey (ACS) from 2010 to 2019 (i.e. 2-3 years of data), however neither of these estimates provide temporal resolution of the IRS data and many of the county estimates are suspect because the ACS surveys do not have a large enough sample in any given year to make estimates below the state level.    

\paragraph{International Migration Data}

Estimation of international migration is complex and important area of research. Recently, Abel and Cohen \cite{abel2019bilateral} updated the migration estimates from Azose and Raftery \cite{azose2019estimation} which cover 200 countries every 5 years from 1990 to 2015. This is the same data used in Abel et. al. \cite{abel2021form} and the estimates we use in this article. The statistical method used to develop these estimates is built on the work by Azose and Raftery \cite{azose2019estimation} which builds on the work by Abel \cite{abel2013estimating,abel2018estimates}. 

This data is developed by first  gathering data on country-level migration stocks, desegregated by country of birth based on administrative and United Nations records. Next, the researchers employ demographic balancing equations to harmonize the data between two periods. The idea being that any change in the migration stock must align with component change in fertility, mortality and migration in a given country. These models use the fertility and mortality information from the United Nations World Populations Prospects to estimate the country-to-country migration flow data at five year intervals. It is worth noting that Abel's \cite{abel2013estimating} original construction more closely follows report data and Azose and Raftery's \cite{azose2019estimation} employ a Bayesian model to produce the final estimates. Abel et al \cite{abel2021form} describe the following major limitations to these migration data is that statistical models do not reconcile migration reports of sending and receiving countries which are often not in agreement. However, these harmonized and estimated migration data are generally considered the best migration data under current standards \cite{abel2019bilateral}.

\paragraph{Spatial-Temporal Co-clustering Model}

We consider three-way data ${\cal X}\in\mathbb{R}^{I\times J\times K}$, where ${\cal X}(i,j,k)$ represents the number of individuals who moved from location $i$ to location $j$ in year $k$.
We expect that migration occurs in systems that arise organically and are largely driven by sociological, economic, and demographic factors, as well as major local events such as a natural disaster. The migration patterns are grouped through the ``origin locations'' (i.e., locations where people move from) and ``destination locations'' (i.e., where people move to), and temporal pattern of this movement. Such a set of data can be represented as in a multi-way tensor co-clustering framework \cite{papalexakis2013from}. Specifically, we model ${\cal X}$ (the origin by destination by time array) as the following decomposition:

\begin{align}
{\cal X} \approx& \sum_{f=1}^F {\bf a}_f\circ{\bf b}_f \circ{\bf c}_f, \numberthis \label{eq:cpd}
\end{align}

where $\circ$ denotes the outer product, i.e., 
\[  [{\bf a}_f\circ {\bf b}_f]_{i,j} ={\bf a}_f(i){\bf b}_f(j),~ [ {\bf a}_f\circ {\bf b}_f \circ {\bf c}_f]_{i,j,k} ={\bf a}_f(i){\bf b}_f(j){\bf c}_f(k). \]
Here,
${\bf a}_f\circ {\bf b}_f\circ {\bf c}_f$ represents the $f$th co-cluster -- a migration system over time in our context. The vector ${\bf a}_f$ indicates the membership (or the degree of association) of the $I$ origin locations with co-cluster $f$. For example, ${\bf a}_f(i)=0$ means that county $i$ is not in the $f$th migration system. The ${\bf b}_f$ vector is defined similarly for the destination locations. Note that ${\bf a}_f\circ{\bf b}_f={\bf a}_f{\bf b}_f^T$ is a rank-one matrix, and defines a bipartite clique (i.e., fully connected bipartite sub-network) over the origin-destination network. The vector ${\bf c}_f$ scales the clique over time, which can be regarded as the clique's temporal signature. Intuitively, ${\bf c}_f(k)$ being a large value means that the $f$th migration system has intense migration activities at the $k$th year.

The model in \eqref{eq:cpd} is the so-called canonical polyadic decomposition (CPD) of third-order tensors, if $F$ is the smallest integer that makes \eqref{eq:cpd} hold exactly. In such cases, $F$ is referred to as the tensor rank.
Finding the CPD expression of the migration data, under our hypothesis, can reveal major migration co-clusters and their activity levels over time.

The key advantage of CP decomposition is that the rank-one components it produces are unique, and can thus be better interpreted. This is to be contrasted with bilinear (matrix) factor analysis methods, which do not produce unique rank-one components. Taking SVD for example and absorbing the singular values into the left and right matrix factors, we can obtain another {\em equivalent} decomposition of the given low-rank matrix. The reason that SVD itself is unique is because we insist on orthogonality of the singular vectors. But the `true' underlying components that we seek in applications are rarely orthogonal, and thus SVD fails to unravel them. 

Owing to the inherent uniqueness of CP  decomposition, we cannot enforce its components to be orthogonal, as the true generating latent factor matrices are not orthogonal. The result is that the variance explained by the sum of CP  components is not the sum of the variances explained by the individual components, so we cannot talk about the variance explained by a single component in isolation, as in SVD. However, we can extract a set of $F$ principal CP components, which together best explain the given data, and because they are unique there is no ambiguity in visualizing them, as there is in the matrix case. 

To be more precise, the co-clustering algorithm tackles the following optimization problem:
\begin{align}\label{eq:formulation}
\minimize_{\mathbf{A}, \mathbf{B}, \mathbf{C}} &\quad \norm{ {\cal W} \circledast\left( \mathcal{X} - \sum_{f=1}^F {\bf a}_f\circ {\bf b}_f\circ{\bf c}_f \right) }_F^2, \\
\st& ~\mathbf{A}\geq {\bf 0}, \mathbf{B}\geq {\bf 0}, \mathbf{C} \geq {\bf 0},   \nonumber 
\end{align}
where ${\bf A}=[{\bf a}_1,\ldots,{\bf a}_F]$, ${\bf B}=[{\bf b}_1,\ldots,{\bf b}_F]$, ${\bf C}=[{\bf c}_1,\ldots,{\bf c}_F]$,  ${\cal W}\in \mathbb{R}^{I\times I\times K}$ is a weight tensor such that
\[ {\cal W}(i,i,k)=0,~\forall i,~\forall k,~~~{\cal W}(i,j,k)=1,~\forall k,~\forall i \neq j, \]
and $\circledast$ denotes the Hadamard product.
The nonnegativity constraints imposed on ${\bf A}$, ${\bf B}$ and ${\bf C}$ reflect their physical interpretations. The weighting discards the diagonal entries in each slab of the data tensor, since entries like ${\cal X}(i,i,k)$ always dominate in terms of magnitude, but they represent static residents in county $i$ and do not encode movements.

\paragraph{Algorithm, Software, and Hyperparameter Selection}
The formulation in \eqref{eq:formulation} entails a special tensor completion problem. Many off-the-shelf algorithms have been designed to handle this problem and its variants; see \cite{papalexakis2013from,fu2020computing,acar2011scalable}.
In this work, we employ the well-optimized and freely available \texttt{Tensorlab} software toolbox \cite{tensorlab3.0} to solve the formulated problem in \eqref{eq:formulation}.
Tensorlab is a Matlab toolbox that is widely used in the signal and data analytics community.
The software has a suite of flexible functions that can deal with plain-vanilla tensor decomposition and tensor decomposition with multiple constraints, e.g., nonnegativity, sparsity, and smoothness. 
The software can also easily handle missing values. In a nutshell, \texttt{tensorlab} treats a wide range of tensor decomposition problems as a nonlinear least squares problem, and recasts these problems into a form that can be dealt with using a Gauss-Newton (GN) nonlinear programming framework. The subproblems in the GN framework are handled using conjugate gradient, which can effectively exploit the multilinear structure of tensor problems to come up with lightweight updates. A tutorial of \texttt{tensorlab}'s basic framework and updating rules can be found in \cite{fu2020computing}.
Users who are not familiar with tensors and nonlinear programming may also use \texttt{tensorlab} as a black-box.

The proposed method has only one hyperparameter to select, namely, the tensor rank of the model, which corresponds to the number of migration systems.
The tensor rank is analogous to the number of principal components in the matrix principal component analysis (PCA) case. For real-life data, due to the existence of noise and modeling error, the data tensors tend to have high (or full) rank. However, the ``useful signal part'' of the tensor is believed to have a low rank due to the high correlations across different modes.
Unlike matrix PCA, incrementally extracting $F$ components from the tensor one by one does not ensure that one will extract the $F$ best (most significant) rank-one components from the data---due to the lack of orthogonality of the latent factors. Furthermore, even extracting the principal CP component is NP-hard in general, see \cite{sidiropoulos2017tensor} and references therein. Nevertheless, we do have good software tools such as \texttt{tensorlab} that work very well in practice, and when the latent factors ${\bf A}$, ${\bf B}$ and ${\bf C}$ are nonnegative and sparse, then even incremental extraction often produces the most prominent $F$ components, as observed in \cite{papalexakis2013from}. In our case, the latent factors are indeed nonnegative and sparse, and thus we have good reason to believe that the $F=6$ migration systems extracted from both datasets are the most prominent ones. In the SI appendix, we present evidence supporting our choice of this single hyperparameter, i.e., setting $F=6$. It turns out that further increasing $F$ does not change the first 6 communities significantly, which validates our postulate. 


\subsection*{Related Works} 

The co-clustering idea was first introduced in \cite{papalexakis2013from} for discovering communities from email networks over time. 
Tensor-based co-clustering was also found useful in analytical chemistry \cite{bro2012coclustering}. 
Variations of tensor co-clustering were recently used for football team clustering, Wikipedia user clustering, and autonomous systems analysis \cite{gujral2020beyond}.
In terms of migration data analysis,
a short workshop paper presented preliminary results of using tensor models to discover the most significant migration clique spatio-temporal migration data. There, instead of using optimization-based low-rank decomposition as in our work, a Bayesian inference framework was used, where the migration counts were modeled as drawn from Poisson distributions and the factor matrices were given Gamma priors \cite{nguyen2017understanding}. The Bayesian nature of the work in \cite{nguyen2017understanding} may make the method heavily dependent on priors, which are not known for real-world data. Non-parametric approaches and those that use as few assumptions and parameters as possible are preferable for exploratory analysis.

\subsection*{Comparison with Dynamic Approaches to Community Detection}

\paragraph{Comparison methods: Walktrap}
To detect any changes in the migration pattern before and after Hurricane Katrina, we construct an aggregated pre-Katrina migration matrix and an aggregated post-Katrina migration matrix.  More precisely, let $\mathbf W_{\text{pre}} =\sum_{n=1990}^{2004}\mathbf W_i-\text{diag}\left(\sum_{n=1990}^{2004}\mathbf W_i\right)$ denote the  aggregated pre-Katrina  weight matrix, where $\mathbf W_i$ denotes the weight matrix associated with the $i$-th observation period  and $\text{diag}\left(\sum_{n=1990}^{2004}\mathbf W_i\right)$ is a diagonal matrix that holds the diagonal elements of $\sum_{n=1990}^{2004}\mathbf W_i$ on its diagonal. Note, that we do not have details on migration patterns within the county, thus the diagonal elements of $\mathbf W_{\text{pre}}$ are set to zero. We construct the aggregated post-Katrina migration matrix ($\mathbf W_{\text{post}} =\sum_{n=2006}^{2018}\mathbf W_i-\text{diag}\left(\sum_{n=2006}^{2018}\mathbf W_i\right)$) analogously. The clustering method   Walk Trap proposed in  \cite{pons2005computing} and applied on  $\mathbf W_{\text{pre}}$  and $\mathbf W_{\text{post}}$  yields the communities depicted in Figure \ref{fig:Pre-Post-Katrina-RandonWalk}.  The community containing New Orleans is depicted in Figure  \ref{fig:Pre-Post-Katrina-NewOrleans}.

\paragraph{Comparison method: InfoMAP} The InfoMAP \cite{rosvall2008maps} community detection method is an information theoretic-based community detection technique. 
It receives an adjacency matrix that represents a directed and weighted network (e.g., our international migration data), and produces a list of hard-clustering (i.e., full partitioning) communities with the goal of optimally compressing information flow. InfoMAP is directly applied to the international migration data, for each period. The corresponding results for the 5 different periods are shown in Figure~\ref{fig:rosvall2008maps_full}. In addition, to reflect changes in community structure over time, we focus on the communities including either the US or China, as these are likely the most important members. The results are depicted in Figure~\ref{fig:rosvall2008maps_small}.

\section*{Supporting Information Appendix (SI)}

All supporting information in the SI Appendix.

\subsubsection*{SI Data sets} 
All data and code will be deposited in Harvard Dataverse (URL/DOI TBD) if accepted for publication. 


\section*{Acknowledgements}

Zack W. Almquist, Tri Duc Nguyen, Mikael Sorensen, Xiao Fu, Nicholas D. Sidiropoulos recieved partial support for this research through ARO Award \#W911NF-19-1-0407. Almquist also received partial support for this research from a Eunice Kennedy Shriver National Institute of Child Health and Human Development research infrastructure grant, P2C HD042828, to the Center for Studies in Demography \& Ecology at the University of Washington.

\bibliography{sa_bib}
\bibliographystyle{Science}

\end{document}